\documentclass [12pt]{article}
\usepackage{lscape}                                           %
\usepackage[centertags]{amsmath}
\usepackage{amsfonts} \usepackage{amssymb}\usepackage{eufrak}
\usepackage{graphicx}
\usepackage{a4}
\usepackage{cite}
\usepackage{lscape}

\begin{document}
\begin{titlepage}
\rightline{}
\rightline{}
\vskip 3.0cm
\centerline{\LARGE \bf  More About  Branes }
\vskip 0.5cm
\centerline{\LARGE \bf on a General Magnetized Torus}
\vskip .5cm
\vskip 1.0cm \centerline{\bf L. De Angelis$^a$, R. Marotta$^b$, F. Pezzella$^b$ and  R. Troise$^a$\footnote{XXIII  Ciclo di Dottorato in Fisica Fondamentale e Applicata}}
\vskip .6cm \centerline{\sl  $^a$ Dipartimento di
Scienze Fisiche, Universit\`a degli Studi ``Federico II'' di Napoli}
\centerline{\sl Complesso Universitario Monte
S. Angelo  ed. 6, via Cintia,  80126 Napoli, Italy}
\vskip .4cm
\centerline{\sl $^b$ Istituto Nazionale di Fisica Nucleare, Sezione di Napoli}
\centerline{ \sl Complesso Universitario Monte
S. Angelo ed. 6, via Cintia,  80126 Napoli, Italy}
\vskip 0.4cm

\begin{abstract}
In the framework of low-energy effective actions of branes compactified on magnetized extra-dimensions, we determine Yukawa couplings for the chiral matter described by open strings attached to D9 branes having different oblique magnetization and living on a general torus $T^{6}$ with an arbitrary complex structure. These results generalize the ones existing in literature.
\end{abstract}

\end{titlepage}

\newpage

\tableofcontents
\vskip 1cm

\section{Introduction}

String Theory provides  a consistent quantum description of all the interactions
including gravity.  The  natural presence of gauge groups in Heterotic and Type I models has
generated, since the origins of the theory, a huge interest in the so-called String Phenomenology, i.e. in exploring the connection between
String and Standard Model physics  (see, for example, ref. \cite{IU}).   In Type II theories, a fundamental role in this respect is played by D-branes.
These solitonic objects define open string degrees of freedom on their world-volume and
can be engineered so that the massless excitations of
such open strings reproduce the Standard Model gauge group (for recent reviews, see for example \cite{Review1,Review2,Review3}).
In compact spaces, configurations of branes ``dressed'' with  constant
magnetic fields along the compact directions of their world-volume - hence the name of {\em magnetized branes} - introduce chiral matter \cite{magnetized1,magnetized2,magnetized3,magnetized4}.   This is the reason why magnetized brane configurations are promoted to new string
theory vacua possibly containing the Standard Model and/or its supersymmetric extensions.
In particular,
even
 neglecting the dynamics of gravity, they provide a tool to study the dependence of the four-dimensional effective field theory actions on the details of compactification both to the tree \cite{Review1,Review2,Review3,0512067,1201.3604}  and to the first order of the perturbative expansion \cite{0508043,1112.5156}. In this paper, such program is realized  in a peculiar
framework of toroidal compactification of ten-dimensional models and  with a particular interest in Yukawa couplings.

More specifically, a stack of $M$ D9-branes is considered in the compact
background $T^6$, being the torus a six-dimensional complex manifold
with a completely arbitrary complex structure. In the same spirit of ref.
\cite{0404229} (see also \cite{also1,also2,also3,also4,also5,also6}), we turn on, along the compact directions, constant
magnetic fields in the abelian sector of the $U(M)$ gauge group
defined on the world-volume of the $M$  branes. Depending on the choice of such constant
fields, the single stack of branes is now separated in  different piles
of magnetized branes.
The ten-dimensional ${\cal N}=1$ super Yang-Mills theory supported in the world-volume of a
stack of D9-branes is dimensionally reduced to four dimensions by
expanding the ten-dimensional bosonic or fermionic fields in a basis of
eigenfunctions of the internal Laplace or Dirac operator.

Two sectors of open strings appear. One corresponds to {\em dy-pole} strings, i.e.  open strings with both ends on the same brane \cite{0709.4149, 0906.3033}, containing fields in the adjoint representation of the gauge group, while the other corresponds to  {\em dy-charged} strings, i.e. open strings ending on two piles of magnetized branes with
different magnetization and describing chiral matter.
In this paper, the analysis  is restricted to the latter sector. The eigenfunctions for the matter fields  have to be
invariant, up to gauge transformations, when translated along the one-cycles of the torus. They are easily determined in the complex frame
where both the metric and the difference $ F^{ab}=F^a-F^b$ of the magnetic fields on the
two piles $a$ and $b$ of branes between which  the strings are stretched,  are diagonal matrices in their off-diagonal boxes. In this frame,
the supersymmetry has been also partially imposed by requiring the field
$F^{ab}$ to be  a $(1,1)$ form in the coordinate system defining the complex torus.
 The wave function is obtained by solving the internal Laplace-Beltrami or Dirac
equation, depending on the Bose-Fermi statistic
of the fields,  with suitable boundary conditions dictated by the torus geometry
 and by the presence of the background magnetic field.
By introducing a suitable ansatz for such a solution, the  lattice identification due to the magnetized
torus geometry reduces to the quasi-periodicity conditions
satisfied by the Riemann Theta function only when the background gauge field, in the original
system of coordinates defining the torus, is a matrix with null diagonal blocks.
The Riemann Theta function turns  out to be dependent on a  generalized complex structure with entries related
to the original complex structure of the torus - or to  its complex
conjugate - which depends on the signs of the eigenvalues of the non-vanishing blocks of the gauge field $F^{ab}$.
In more general
configurations, it seems that the Theta function is no longer the solution
of the internal wave operator, even if a final answer to this problem
requires a deeper analysis.

The coefficients of the dimensionally reduced four-dimensional  effective
action are obtained by evaluating overlap integrals over different wave
functions. The Yukawa couplings among two fermions and one scalar field
are obtained by evaluating  an  integral over  three of such functions,
each of them defined in the local systems of the coordinates where the
metric and the gauge field take the aforementioned simple form.
In order to compute that integral,  all the wave
functions have been rewritten in the common system of real coordinates where the torus is
trivially defined. Analogously with the corresponding calculus done
in the case of the factorized torus, an identity between
the product of two Riemann Theta functions has been used.  It is derived in
ref. \cite{0904.0910} generalizing  some results given
in ref. \cite{MumfordI}. In getting that identity,
an arbitrary matrix $\alpha$ has been introduced. It is restricted only by the necessity to make  the matrix $\alpha
(I_{ab}^{-1}+I_{bc}^{-1})^{-1}$ have integer entries, being $I_{ab}$ and $I_{bc}$ the first Chern
classes associated with the corresponding differences of the gauge fields
on the three branes labeled by $a,b$ and $c$. The first Chern classes are involved in the
calculus of the Yukawa couplings.
The simplest  choice $\alpha={\rm det}  \left[ I_{ab}I_{bc} \right] \mathbb{I}$ \cite{0904.0910} allows one to
evaluate the integral, getting a linear combination of
Rieman Theta functions as a result. This expression is in agreement with the one
obtained in ref. \cite{0904.0910} in the case of the torus $T^4$ with
trivial complex structure and extends it to the case of the torus $T^6$ with arbitrary complex structure.

When all of  the magnetic fields which are active on the three stacks of branes are independent but commute, it is possible to explicitly evaluate the sum over the
Riemann Theta functions, obtaining an expression for the Yukawa couplings compatible  with the analogous stringy result given in ref. \cite{0709.1805}. The same result is obtained by taking
$\alpha=I_{ab}I_{bc}$, which, in the case of commuting first Chern classes,
trivially makes
$\alpha (I_{ab}^{-1}+I_{bc}^{-1})^{-1}$ an integer matrix. This coincidence enforces the
validity of the result here obtained.

The paper is organized as follows.

In sect. 2, generalities about dimensional reduction and magnetic fluxes are given. In particular, some helpful complex
coordinates are discussed in which both the metric and the magnetic fluxes become diagonal matrices in the non-vanishing blocks. In sect. 3, it is first shown how the effective four-dimensional action of the dy-charged strings  can be derived from the ten-dimensional super Yang-Mills action with gauge group $U(M)$ through a Kaluza-Klein reduction. Then,  the bosonic and fermionic KK mass spectra are obtained. For the lowest components of these fields, the internal wave-function has been obtained and compared with the analogous results in literature. In sect. 4, Yukawa couplings for a general magnetized torus $T^{6}$ have been computed. This has been done in the case of both arbitrary and commuting first Chern classes. Finally, the five appendices contain many explicit details of computation.  In particular, in appendix A  notations on space-time indices are fixed;  in appendix B one can find generalities on the torus $T^{2d}$ and a discussion on some possible choices of coordinates; in appendix C relevant properties of the Riemann Theta function are discussed  and details on  the derivation of  the bosonic and fermionic wave-functions are given;  appendix  D contains explicit calculations of the Yukawa couplings and appendix E deals with the main properties of a string in a magnetic background.

\section{Dimensional reduction and fluxes}
\label{reduction and fluxes}

A configuration made of a stack of $M$ D9-branes in the compact background ${T^{2d}}$ is going to be studied in this paper, mainly in the case $d=3$. Branes backreaction on the space-time  geometry is neglected and our analysis is focused on the open string degrees of freedom. Their interaction with the closed string degrees of freedom is described by the supersymmetric  DBI and by Chern-Simons actions. In the following, attention will be driven to the low-energy limit of the DBI action which turns out to be, for this particular brane configuration, the ten-dimensional ${\cal N}=1$ super Yang-Mills with gauge group $U(M)$. This theory does not contain chiral matter, which is a fundamental ingredient of the Standard Model and/or of its supersymmetric extensions. Here chiral matter can be  introduced by turning on some abelian constant background magnetic fields \cite{BachasPorrati,0404229} along the compact dimensions of  the world-volume of $N_a$ branes, with $\sum_{a=1}^n N_a=M$. In this way,  the four-dimensional Lorentz invariance is preserved.   The integer $n$ gives the number of piles  of magnetized branes having different magnetizations. The original gauge group is then broken into the product of $U(M)\sim \prod_{a=1}^n U(N_a)$ and the chiral or dy-charged matter is given by the open strings ending on two  stacks, $N_a$ and $N_b$, of branes with different magnetization. It  belongs to the bifundamental representation $(N_a,\,\bar{N}_b)$ of the gauge group $U(N_a)\times  U(N_b)$.

The quantity playing an important role in the forthcoming study is the difference between the background magnetic fields active on the world-volume of the two piles  $a$ and $b$:
\begin{eqnarray}
F^{ab} & \equiv &  \frac{1}{2}( F^a-F^b)_{MN} dX^{M} \wedge dX^{N}    \nonumber \\
&   =  & \frac{1}{2} {F^{(xx)ab}_{mn}} dx^m\wedge dx^n + F^{(xy)ab}_{mn} dx^m\wedge dy^n +\frac{1}{2} F^{(yy)ab}_{mn} dy^m \wedge dy^n  \label{1}
\end{eqnarray}
$(M,N=1, \dots, 2d)$, where the curved space-time coordinates $(x^{m}, y^{m}) \equiv (X^{2m-1}, X^{2m})$ with $m=1, \dots, d$ are identified by the torus geometry:
\begin{eqnarray}
x^m\equiv x^m+2\pi Rm_{1}^{m} \quad;\quad y^m\equiv y^m+2\pi R m_{2}^{m}   \qquad \vec{m}_{1}, \vec{m}_{2} \in \mathbb Z^{d}  \nonumber
\end{eqnarray}
being $R$ an arbitrary dimensional  parameter introduced to deal with the dimensionful $(x^{m}, y^{m})$.

The field strength  $F^a$ carries also gauge indices corresponding to the $U(1)$ subgroup of the  gauge group $U(N_a)$ and the associated fluxes thread the 2-cycle $(M,N)$  of the torus.
The  components of $F^a$ can be expressed as:
\begin{eqnarray}
F^a_{MN}=\frac{1}{2\pi R^{2}}I_{MN}^{a}\frac{\mathbb{I}_{N_a}}{N_a} \nonumber
\end{eqnarray}
being $I^{a}_{MN}$ the first Chern class,
defined as the flux of $F^{a}$ through the 2-cycle $(M,N)$ of the torus:
\begin{eqnarray}
I^{a}_{MN} = \int_{(M,N)}{\rm Tr}\left[ \frac{F^a}{2 \pi}\right]  \,\, .
\end{eqnarray}
from which the antisymmetry of $I^{a}_{MN}$ in $M$ and $N$ follows.

In the following, only the case $N_a=1$, for each $a$, is going to be considered, which corresponds to  the complete breaking of the gauge group $U(M)\sim U(1)^M$.  The breaking of the original gauge group $U(M)\sim\prod_{a=1}^n U(N_a)$ is straightforward \cite{0404229}.

The torus can be seen as a complex manifold by introducing the curved complex coordinates
\begin{eqnarray}
w^m=\frac{x^m+U^m_{~n}y^n}{2\pi R}\quad;\quad \bar{w}^m=\frac{x^m+\bar{U}^m_{~n}y^n}{2\pi R} \nonumber
\end{eqnarray}
with the identifications
\begin{eqnarray}
w^m & \equiv  & w^m+ m_{1}^{m}+ U^m_{~n}m_2^{n}  \nonumber \\
\bar{w}^m & \equiv &  \bar{w}^m+ m_{1}^{m}+ \bar{U}^m_{~n}m_2^{n} \,\,    \,\, .\nonumber
\end{eqnarray}
Here  $U$  is a complex matrix parametrizing the complex structure of the torus.
In the system of complex coordinates,  the gauge field takes the following form:
\begin{equation}
F^{ab}=-\frac{(2\pi R)^2}{8}  F^{{(\cal{WW}}) ab}_{MN} d{{\cal{W}}^M} \wedge   d{{\cal{W}}^N} \nonumber
\end{equation}
with $( {\cal W}^{1}, \dots, {\cal W}^{2d}) \equiv (w^1, \dots, w^d, \bar{w}^1, \dots \bar{w}^d)$ and
\begin{eqnarray}
&&F^{(ww)}= ({\rm Im}U^{-1})^t \left[\bar{U}^t \,F^{(xx)} \bar{U} -\bar{U}^t F^{(xy)} +{F^{(xy)t}}\bar{U} +F^{(yy)}\right] {\rm Im}U^{-1} \nonumber\\
&&{ F}^{(w\bar{w})}= ({\rm Im}U^{-1})^t \left[-\bar{U}^t F^{(xx)} U+\bar{U}^t F^{(xy)} -{F^{(xy)t}}U-F^{(yy)}\right] {\rm Im}U^{-1} \label{108}
\end{eqnarray}
while ${F}^{(\bar{w}w)}= F^{(w\bar{w})*}$ and ${F}^{(ww)}=F^{(\bar{w}\bar{w})*}$.\footnote{In this analysis the $a$,$b$ labels are omitted when possible.} Supersymmetric configurations require the gauge field to be a $(1,1)$ two-form, i.e. ${F}^{(w w)}~=~{F}^{(\bar{w}\bar{w})}=0$ \cite{GSW}. By separately imposing these two constraints on the non-vanishing components of the gauge field, the following identity is derived:
\begin{eqnarray}
\! \! \! \! \! {F}^{(w\bar{w})}= -2\,i{\rm Im}U^{-t}(\bar{U}^tF^{(xx)}+{F^{(xy)t}})=-2\,i  ({F^{(xx)t}}U+F^{(xy)}){\rm Im}U^{-1} \label{15}
\end{eqnarray}
where $t$ denotes matrix transposition and which shows that $iF^{(w \bar{w})}$ is an Hermitian matrix \cite{0904.0910}.
It is convenient to rewrite the magnetic flux in the system of flat coordinates defined in eq. (\ref{flcomp}) of the appendix \ref{torus}. In this frame $i{F}_{r\bar{s}}^{(w\bar{w})ab}=i e^m_{~r} F^{(w \bar{w}) ab}_{m \bar{n}}\bar{e}^{\bar{n}}_{~\bar{s}}$ is still hermitian and  can be diagonalized by an  unitary matrix:
\begin{eqnarray}
F_{r \bar{s}}^{(w \bar{w})ab} \, (\bar{C}_{ab}^{-1})^{\bar{s}}_{~\bar{p}} = \frac{2}{i} \frac{\lambda_p^{ab}}{(2\pi R)^2}\delta_{r\bar{s}} \,(\bar{C}_{ab}^{-1})^{\bar{s}}_{~\bar{p}}\label{eqei}
\end{eqnarray}
with $(\bar{C}_{ab}^{-1})^{\dagger} \bar{C}_{ab}^{-1}=\mathbb{I}$ and where $r,\bar{s}=1, \dots, d$. This condition can also be seen as the orthonormality condition of the eigenvectors of the hermitian matrix.
In the system of curved complex coordinates, by introducing the matrices $(\bar{C}_{ab}^{-1})^{{\bar{n}}}_{~\bar{s}} =\bar{e}^{\bar{n}}_{~\bar{r}}(\bar{C}_{ab}^{-1})^{{\bar{r}}}_{~\bar{s}}$ and using the identity  $(C^{-1}_{ab})^m_{~r} h_{m\bar{n}} (\bar{C}_{ab}^{-1})^{{\bar{n}}}_{~\bar{s}}=\delta_{rs}$ where $h_{m\bar{n}}$ refers to the metric of the complex torus and
which is a direct consequence of the unitarity  condition of the $\bar{C}_{ab}^{-1}$ matrices, one can write:
\begin{eqnarray}
(C_{ab}^{-1})^m_{~r}\,\, F_{m \bar{n}}^{(w, \bar{w})ab} \,\, (\bar{C}_{ab}^{-1})^{{\bar{n}}}_{~\bar{s}} = \frac{2}{i} \frac{\lambda_r^{ab}}{(2\pi R)^2} \delta_{r\bar{s}} \,\, . \nonumber
\end{eqnarray}
 This diagonalization naturally defines a new  system of complex coordinates:\footnote{We could give an equivalent definition of  this new frame by starting from the curved complex coordinates and writing $w^m= \left(C^{-1}_{ab}\right)^m_{~s}{z}^s$ and $\bar{w}^m= \left(\bar{C}_{ab}^{-1}\right)^{\bar m}_{~\bar{s}}\bar{z}^s$.}
\begin{eqnarray}
w^s= \left(C^{-1}_{ab}\right)^s_{~r}{z}^r_{ab}~~;~~\bar{w}^{s}= \left(\bar{C}^{-1}_{ab}\right)^{{\bar{s}}}_{~\bar{r}}\bar{ z}^{r}_{ab}\label{19}
\end{eqnarray}
with the metric given by:
\begin{eqnarray}
ds^2=
(2\pi R)^2\delta_{s \bar{r}} d{z}^s_{ab} d\bar{z}^r_{ab}\label{metr}
\end{eqnarray}
as follows from the orthonormality conditions of  the eigenvectors $C^{-1}_{ab}$. These are local, which means that they depend on the magnetic fields that we are diagonalizing and, therefore,  on the dy-charged sector of the open string under consideration. The lattice identification is now given by
\begin{eqnarray}
{z}_{ab}^r\equiv {z}_{ab}^r +  (C_{ab})^r_{~m} \, {m_1}^m+   (C_{ab})^r_{~m} \, U^m_{~n} \, {m_2}^n\label{id1}
\end{eqnarray}
and, after defining $({\cal Z}_{ab}^1,\dots, {\cal Z}_{ab}^{2d})=({z}_{ab}^{1}, \dots, {z}_{ab}^{d}, \,{\bar z}_{ab}^1,\dots,  {\bar z}_{ab}^d )$, the metric is read from:
\[
\frac{ds^2}{(2\pi R)^2} =\frac{1}{2}\, d{\cal Z}^I_{ab}\, {\cal G}_{IJ} \,  d{\cal Z}_{ab}^J
\]
with
\begin{eqnarray}
{\cal G}=\left( \begin{array}{cc}
0&\mathbb{I}\\
\mathbb{I}&0\end{array}\right)\label{cmpxmtr}
\end{eqnarray}
while the  magnetic field strength is
\begin{eqnarray}
F^{ab}
&=& \frac{1}{2} F^{({\cal Z} {\cal Z}) ab}_{IJ} d{\cal Z}_{ab}^I\wedge d{\cal Z}_{ab}^J~~;~~F_{IJ}^{({\cal Z} {\cal Z}) ab}= \frac{i}{2}\left(\begin{array}{cc}
0& {\cal{I}}_{\lambda}^{ab}\\
-{\cal{I}}_{\lambda}^{ab}&0\end{array}\right)\label{89}
\end{eqnarray}
being
\begin{eqnarray}
{\cal{I}}_{\lambda}^{ab}={\rm diag} \left( \lambda_1^{ab}  \dots  \lambda_{d}^{ab} \right)
\end{eqnarray}
 and $I,J=1, \dots, 2d$ flat indices.
We notice that in the ${\cal Z}_{ab}$-coordinates  both the metric and the field strengths are diagonal matrices in the non-vanishing blocks and this feature will be the key ingredient in finding the solutions of the Laplace and Dirac equations associated with the compact directions.

\section{The four-dimensional effective action}
Following the procedure defined in ref. \cite{0404229} and here summarized (see also \cite{0810.5509} for a better understanding of the notations used in this paper), the four-dimensional action is obtained starting from the ten-dimensional super Yang-Mills action with gauge group $U(M)$:
\begin{eqnarray}
S = \frac{1}{g^2} \int d^{10} X^{ \hat{N}} \,\, {\mbox Tr}
\bigg(- \frac{ 1}{ 4} {\cal F}_{\hat{M}\hat{N}} {\cal F}^{\hat{M}\hat{N}}  +
\frac{i}{  2}\bar{\lambda} \Gamma^{\hat{M}} {D}_{\hat{M}} \lambda \bigg)
\label{10dimla}
\end{eqnarray}
where $\hat{M}, \hat{N} = 0,\dots,9$, \, $g^2 =  4 \pi {\rm e}^{\phi_{10}} (2 \pi \sqrt{\alpha'})^6$ and
\begin{eqnarray}
{\cal F}_{\hat{M}\hat{N}} = \nabla_{\hat{M}} A_{\hat{N}} - \nabla_{\hat{N}} A_{\hat{M}}  - i[A_{\hat{M}}, A_{\hat{N}}]~~;~~
{D}_{\hat{M}} \lambda = \nabla_{\hat{M}} \lambda - i [A_{\hat{M}},\lambda]
\label{FMN}
\end{eqnarray}
with $ \lambda$ being  a ten-dimensional Weyl-Majorana spinor. The gauge breaking is realized by first separating the generators $U_a$ of the Cartan subalgebra from the ones out of it,  $e_{ab}$, in the definitions of the gauge field and of the gaugino:
\begin{eqnarray}
A_{\hat{M}} = B_{\hat{M}} + W_{\hat{M}} = B_{\hat{M}}^a U_a + W_{\hat{M}}^{ab} e_{ab}~~;~~
\lambda  =  \chi + \Psi = \chi^{a}U_a + \Psi^{ab}e_{ab}
\label{exp98}
\end{eqnarray}
and later expanding the Lagrangian around the background fields which are present only along the compact directions in $T^{6}$ of the branes:
\begin{eqnarray}
B_M^{a}(x^{\mu} , X^{N} ) & = & \langle B_M^{a} \rangle (X^N)+
 \delta B_M^{a}(x^{\mu} , X^N) \nonumber \\
W_M^{ab}(x^{\mu} , X^{N}) & = & 0 + \Phi_M^{ab}(x^{\mu}, X^N) \,\, .
\label{expa3}
\end{eqnarray}
Here $\mu = 0, \dots, 3$ and $M, N=1, \dots, 6$. The fields $B^{a}_M$ and $\Phi_M^{ab}$ are, respectively, adjoint and chiral scalars, from the  point of view of the four-dimensional Lorentz group.  The background fields $\langle B^{a}_{M}\rangle $ are taken with a constant field strength corresponding to the background constant magnetic fields along the compact dimensions, as discussed in the previous section. In particular, the gauge
\begin{eqnarray}
\langle B^{a}_{M} \rangle (X^{N}) = - \frac{1}{2} F_{MN}^{a} X^{N}\,\,  \nonumber
\end{eqnarray}
is chosen.

The following assumes the entire action in terms of  the fields
introduced above.  Only  the relevant terms will be analyzed, namely the quadratic terms involving the scalar and fermion fields and the trilinear terms involving a scalar and two fermions
from which the Yukawa couplings can be computed.

\subsection{The scalar kinetic action}
The quadratic terms in the scalar fields  of the four-dimensional action, derived in detail in
ref. \cite{0404229}, are  obtained by starting from  eq. (\ref{10dimla}) and expanding  the fields  defined in the second line of eq. (\ref{expa3})  in a basis of eigenfunctions of the internal Laplace-Beltrami operator:
\begin{eqnarray}
- \tilde{D}_N\tilde{D}^N \phi_{\cal M}^{ab} (X^N)=m_{{\cal M}}^2\phi^{ab}_{{\cal M}}(X^N) ~~;~~ \Phi_M^{ab}= \sum_{\cal M} \varphi^{ab}_{M, \,{\cal M}}(x^\mu) \otimes \phi^{ab}_{\cal M}(X^N)   \nonumber
\end{eqnarray}
with suitable boundary conditions determined by the torus geometry. Here, the
 covariant derivative depends only on the constant background gauge fields
\begin{eqnarray}
\tilde{D}_N\phi^{ab}_{\cal M}=\partial_N\phi^{ab}_{\cal M}-i  (\langle B_{N}^{a} \rangle-\langle B_{N}^b \rangle)\phi^{ab}_{\cal M}\,\, .  \nonumber
\end{eqnarray}
The resulting action for the lowest excitations of the Kaluza-Klein tower  is the sum of two terms in
 \begin{eqnarray}
S_2^{(\phi)}= \int d^4 x \sqrt{G_4} \,\,  {\cal K} \,\,
\varphi^{N\,\,ba}_{\,\, 0}(x^\mu)   \left[ \left(G^{~M}_{N}  D_{\mu}  D^{\mu}      - \frac{ (M^{2}_{0} )
^{~M}_{N}}{(2 \pi R)^{2}} \right)\varphi_{M,\, 0}  \right]^{ab}
\label{la62}
\end{eqnarray}
 with
\begin{eqnarray}
{\cal K}=\frac{1}{2 g^2} \int d^6 X^{N}
\sqrt{G_6}\phi_{0}^{ba }   \phi^{ab}_{0}~~;~~\frac{ [ ({M}^{2}_{\cal M})]_{N}^{\,\, M} ]^{ab} }{(2 \pi R)^{2}}=m^{2ab}_{\cal M} G_{N}^{M} - 2 i F^{Mab}_{N}  \,\, . \label{bac}
\end{eqnarray}
The Lagrangian will be rewritten in the system of complex coordinates $ {\cal Z}$ in which the mass operator becomes the diagonal matrix
\[
\left[ M^{2}_{\cal M} \right]^{ab}  \equiv \mbox{ diag}~(\tilde{m}_{\cal M}^{2ab }~ \mathbb{I}~- ~2~ {\cal I}_{\lambda}^{ab}, \tilde{m}_{\cal M}^{2ab} \mathbb{I} +~2~ {\cal I}_{\lambda}^{ab} )
\]
with $\tilde{m}_{\cal M}^{ab}= 2\pi R \,\, m_{\cal M}^{ab}$.

In the following we are going to omit the indices $ab$, even if it is clear that we are examining this sector.

The eigenvalues of the Laplace-Beltrami operator are easily determined in this frame. The commutation relations $[\tilde{D}_{I}^{{(\cal Z)}},\,\tilde{D}_{J}^{{(\cal Z})}]=-i\,F_{IJ}$  of the covariant derivatives reduce to the algebra of six decoupled creation and annihilation bosonic operators. In fact, what has been done above so far is valid for any number $d$ of compactified space dimensions. In this general case one therefore has the algebra of $d$ decoupled creation and annihilation bosonic operators.  This is due to the block diagonal expression of the background gauge field. The identification of such operators with the covariant derivatives depends on the signs of the eigenvalues $\lambda_r$, being for positive $\lambda_r$:
\begin{eqnarray}
a_r^\dagger = \sqrt{\frac{2}{|\lambda_r|}}i\,\tilde{D}_r^{({\cal Z})}~~;~~a_r= \sqrt{\frac{2}{|\lambda_r|}} i \tilde{D}_{r +d}^{({\cal Z})} \label{osc0}
\end{eqnarray}
with the role of the creation and annihilation operators exchanged for negative $\lambda_r$. In both cases one has $[a_r,\,a^\dagger_r]=1$ and the Laplace equation becomes
\begin{eqnarray}
\sum_{r=1}^d |\lambda_r| (2N_{r} + 1) \phi_{\cal M} =\tilde{m}^2_{\cal{M}} \phi_{\cal M}~~;~~N_r=a^\dagger_r\, a_r \,\, .   \nonumber
\end{eqnarray}
The eigenvalues of the mass operator result to be:
\begin{equation}
M^{2}_{\pm ; s} = \sum_{r=1}^d  |\lambda_r| (2\, N_r+1) \mp 2 \lambda_{s}  \label{m2}
\end{equation}
where the signs $\pm$ refer respectively to the fields $\varphi_{z^{s},\,{\cal M}}$ and $\varphi_{\bar{z}^{s},\,{\cal M}}$. In these notations one has: ${\cal M}\equiv (N_1, \dots, N_d)$.

According to eq. (\ref{m2}), the lightest states turn out to be $\varphi_{z^r,\,0}$ for positive eigenvalues $\lambda_r$,  otherwise  $\varphi_{\bar{z}^r,\,0}$. In the case $d=3$, among these, the lightest one is massless if the ${\cal N}=1$ susy condition $|\lambda_r|+|\lambda_s|=|\lambda_t|$ ($r\neq s\neq t$) is imposed. Then, by applying  creation operators on the massless state, two towers of Kaluza-Klein states are generated.  Their  spectrum, when the ${\cal N}=1$  susy condition is imposed, is contained in the expression:
\begin{eqnarray}
M^2_k=  2 \sum_{r=1}^{3} |\lambda_r|(N_r+k)~~;~~k=0,\,1\,\, .  \nonumber
\end{eqnarray}
 The same KK spectrum will be shown in a while to be obtained by solving the equation of motion for the fermions. This property is the standard  Bose-Fermi degeneracy that is peculiar of supersymmetric theories.

The eigenfunctions relative to the ground state  is obtained by solving the first-order differential equations
\begin{eqnarray}
a_r \phi_0=0 ~~\forall r \Leftrightarrow \left\{\begin{array}{ll}
                                                        \left(\frac{\partial}{\partial \bar{ z}^r_{ab}}+\frac{i}{2}  F^{( {\cal Z} {\cal Z})}_{r+d\,r} { z}^r \right)\phi_{0}=0& {\rm if}~\lambda_r>0\\
                                                        \left(\frac{\partial}{\partial { z}^r}-\frac{i}{2}  F^{( {\cal Z} {\cal Z})}_{r+d\,r} \bar{ z}^r \right)\phi_{0}=0& {\rm if}~\lambda_r  < 0
                                                        \end{array}\right. \,\, . \label{egs}
\end{eqnarray}

These sets of equations, dependent  on the sign of $\lambda_{r}$, are unified  by introducing new coordinates
$(Z^1, \dots, Z^{2d}) \equiv (\mbox{z}^{1}, \dots, \mbox{z}^{d},  \bar{\mbox{z}}^1, \dots  \bar{\mbox{z}}^d)$ with
\begin{eqnarray}
\mbox{z}^r=z^r~~ \mbox{for}~\lambda_r>0\qquad; \qquad  \mbox{z}^r=\bar{z}^r~~ \mbox{for}~\lambda_r<0  \label{clcoo} .
\end{eqnarray}
In this frame, the background field becomes:
\begin{eqnarray}
F^{( \mbox{z} \mbox{z})}
= \frac{i}{2} \left(\begin{array}{cc}
                    0& |{\cal{I}}_\lambda|\\
                    - |{\cal{I}}_\lambda|&0\end{array}\right)\label{29}
\end{eqnarray}
with the associated vector potential:
\begin{eqnarray}
A_{r}^{(Z)} \equiv \langle B_{r}\rangle = - \frac{1}{2} F_{rs}^{ (\mbox{z} \mbox{z}) }  \bar{\mbox{z}}^s= -\frac{i}{4}|\lambda_r|\bar{\mbox{z}}^r \,\, .  \nonumber
\end{eqnarray}
 Eqs. (\ref{egs}) are now expressed in a unique differential equation
\begin{eqnarray}
\left(\frac{\partial}{\partial \bar{\mbox{z}}^r}+\frac{1}{4} |\lambda_r|\,\mbox{z}^r\right)\phi_0=0\label{kge} \,\, .
\end{eqnarray}
The solution is
\begin{eqnarray}
\phi_0= e^{- \frac{1}{4} \sum_{r=1}^{d} \bar{ \mbox{z}}^r|\lambda_r| \mbox{z}^r} f(\vec{\mbox{z}})=e^{-\frac{1}{4} \vec{\bar{\mbox{z}}}^t\,\,|{\cal I}_\lambda| \vec{\mbox {z}}}f(\vec{\mbox{z}}) \label{fizero}
\end{eqnarray}
with $f(\vec{\mbox{z}})$  being an holomorphic function of the coordinates which is determined by  the boundary conditions.

 It is interesting to notice that the wave-function (\ref{fizero}), when rewritten in the original system of coordinates ${\cal Z}^I=(z^r,\bar{z}^r)$, may depend on both  the holomorphic and anti-holomorphic variables. However, it never simultaneously depends on a variable and its complex conjugate, i.e. on $z^r$ and $\bar{z}^r$ (same $r$).  Therefore, from this point of view, $f$ is always a holomorphic function of the  complex coordinates.

Boundary conditions  are dictated by the transformation properties  of the scalar fields under the torus translations \cite{THooft, 0306006}. The complex torus, in the z-frame, is defined through the lattice identification:
\begin{eqnarray}
\mbox{z}^r\equiv \mbox{z}^r +C^{(\lambda)r}_{~~~~n}\left[ m_1^n+ \Omega^{n}_{~m}m_2^m\right]\label{id2}
\end{eqnarray}
with
\begin{equation}
 \Omega= (C^{(\lambda)})^{-1}  \tilde{C}^{(\lambda)}
\end{equation}
 where
\begin{eqnarray}
{C^{(\lambda)}}^r&=&\left(\frac{1+{\rm sign} (\lambda_r)}{2}\right)
C^r+ \left(\frac{1-{\rm sign} (\lambda_r)}{2}\right)\bar{C}^r=\bar{C}^{(-\lambda)\,r}\nonumber\\
{\tilde C}^{(\lambda)\,r}&=&\left(\frac{1+{\rm sign} (\lambda_r)}{2}\right)
C^r  U+ \left(\frac{1-{\rm sign} (\lambda_r)}{2}\right)\bar{C}^r \bar{U}=\bar{\tilde C}^{(-\lambda)\,r}\label{defcl}
\end{eqnarray}
while $\bar{\Omega}= (C^{(-\lambda)})^{-1} \tilde{C}^{(-\lambda)}$.

We notice that the matrix $\Omega$,  with all the $\lambda_r$s having  the same sign,  coincides with the complex structure of the torus or its complex conjugate. In comparing eq.~ (\ref{id2}) with eq. (\ref{id1}), one can see that it plays the same role as the complex structure $U$ in the $z$-frame. For these reasons $\Omega$ will be named the ``generalized complex structure".

The behavior  of the vector potential $A^{ ( \mbox{z} ) }_{r}$ under the lattice translations
\begin{eqnarray}
A^{ (\mbox{z} )}_{r} (\bar{\mbox{z}}+ \bar{C}^{(\lambda)} \eta_{(s)})\equiv A^{ ({\mbox{z})}}_{r}( \bar{\mbox{z}})+\partial_{r}\chi^{(1)}_{(s)}~~;~~A^{ (\mbox{z} )}_{r}( \bar{\mbox{z}}+ \bar{\tilde{C}}^{(\lambda)} \eta_{(s)})\equiv A^{ (\mbox{z} )}_{r}( \bar{\mbox{z}})+\partial_{r}\chi^{(2)}_{(s)}  \nonumber
\end{eqnarray}
defines the corresponding  gauge transformations
\begin{eqnarray}
\chi_{(s)}^{(1)}&=& -\frac{i}{4} {\mbox{z}}^r|\lambda_r|  (\bar{C}^{(\lambda)})^r_{~n}\eta^n_{(s)}+\frac{i}{4} \bar{\mbox{z}}^r|\lambda_r|  ({C}^{(\lambda)})^r_{~n}\eta^n_{(s)}~~\nonumber\\
\chi_{(s)}^{(2)}&=& -\frac{i}{4} {\mbox{z}}^r|\lambda_r|  (\tilde{\bar{C}}^{(\lambda)})^r_{~n}\eta^n_{(s)}+\frac{i}{4} \bar{\mbox{z}}^r|\lambda_r|  ({\tilde C}^{(\lambda)})^r_{~n}\eta^n_{(s)} \,\, \label{bc}
\end{eqnarray}
with $\eta^t_{(s)}=(\overbrace{0,\dots,0, 1}^{s \,\,\, times},0,\dots)$.
The holomorphic function appearing in the definition of the ground state is determined by imposing the identifications
\begin{eqnarray}
\phi_0(\vec{\mbox z}+ {C^{(\lambda)}} \eta_{(s)},\,\vec{\bar{ \mbox z}}+\bar{C}^{(\lambda)} \eta_{(s)})=e^{i\chi_{(s)}^{(1)}}
\phi_0(\vec{\mbox z},\,\vec{\bar{ \mbox z}})\nonumber \\
\phi_0(\vec{ \mbox z}+ \tilde{C}^{(\lambda)}  \eta_{(s)},\,\vec{\bar{ \mbox z}}+\bar{\tilde C}^{(\lambda)}  \eta_{(s)})=e^{i\chi_{(s)}^{(2)}}
\phi_0(\vec{\mbox  z},\,\vec{\bar{\mbox z}}) .  \,\, \label{bctorus}
\end{eqnarray}
The solution is obtained by writing eq. (\ref{fizero}) in the equivalent form:
\begin{eqnarray}
\phi_0=e^{-\frac{1}{4} \vec{\bar{\mbox  z}}^t  |{\cal I}_{\lambda}|\vec{ \mbox z} +\frac{1}{4}\vec{\mbox z}^t {C^{(\lambda)}}^{-t}\bar{C}^{(\lambda)\,t} |{\cal I}_\lambda| \vec{ \mbox z}}
\theta(\vec{\mbox z})\label{ansatz1}
\end{eqnarray}
and observing that $\theta$ satisfies the following periodicity conditions when translated along the cycles of the hypertorus:
\begin{eqnarray}
\theta( \vec{\mbox z }+ C^{(\lambda)} \eta_{(s)})&=& \theta(\vec{\mbox z})~~\nonumber\\
\theta( \vec{\mbox z}+ C^{(\lambda)} \Omega \eta_{(s)})&=& e^{- \frac{i}{2}\eta_{(s)}^t  {\rm Im}\Omega^t {C^{(\lambda)}}^t {\cal I}_\lambda| \bar{C}^{(\lambda)}[2{C^{(\lambda)}}^{-1}\vec{\mbox z} +\Omega \eta_{(s)}]}
\theta(\vec{\mbox{ z}}) \,\, . \label{IIbc}
\end{eqnarray}
where we have used the identity $ \tilde{C}^{(\lambda)}\eta_{(s)}= C^{(\lambda)}{C^{(\lambda)}}^{-1}\tilde{C}^{(\lambda)}\eta_{(s)}=C^{(\lambda)}\Omega \eta_{(s)}$.

We would like to stress here that the boundary conditions written in eqs. (\ref{IIbc}) are valid if the following condition holds:
\begin{eqnarray}
 F^{(xx)} = 0 \,\, .
\end{eqnarray}
In fact, eqs. (\ref{IIbc})  are derived under the assumption that the product
$C^{(\lambda)t\,}|{\cal I}_\lambda|\,\bar{C}^{(\lambda)}$ is a symmetric matrix. The physical meaning of this constraint comes out if one rewrites the gauge field given in eq. (\ref{29}) in the original  system of real coordinates. Details of this calculation are given in appendix \ref{wave-function}.  The result is that the symmetry of the previous matrix  requires $F^{(xx)}$ to vanish, while one derives:
\begin{eqnarray}
F^{(xy)}= \frac{1}{(2\pi R)^2}{C^{(\lambda)t}}|{\cal I}_{\lambda}|\bar{C}^{(\lambda)}\,{\rm Im}\Omega~~;~~F^{(yy)}= -{F^{(xy)}}^t \Omega+\Omega^t F^{(xy)} \,\, .\label{148}
\end{eqnarray}
By using the previous relations in the second line of eq. (\ref{IIbc}), one gets:
\begin{eqnarray}
\theta( \vec{\mbox z}+ C^{(\lambda)} \Omega \, \eta_{(m)} )= e^{- i\frac{(2\pi R)^2}{2} \eta_{(m)}{F^{(xy)t}} [2{C^{(\lambda)}}^{-1} \vec{\mbox{z}}  +\Omega\,\eta_{(m)}]}
\theta( \vec{\mbox z}) \,\, .\label{fbc}
\end{eqnarray}
The solution of eq.  (\ref{kge}) is characterized by a Riemann Theta function with the boundary conditions in eq. (\ref{fbc}):
\begin{eqnarray}
\theta( \vec{\mbox{z} })\equiv \Theta\left[\begin{array}{c}\vec{j}\\0\end{array}\right]
(I {C^{(\lambda)}}^{-1}\vec{\mbox{z}}|I \Omega)\label{sol}
\end{eqnarray}
with
\begin{eqnarray}
I= (2\pi R)^2\frac{{F^{(xy)t}}}{2\pi} \,\, .
\end{eqnarray}
This is known to be well-defined  only if the matrix  $ {F^{(xy)t}} \Omega$  is  symmetric and
$ {F^{(xy)t}}{\rm Im} \Omega>~0$ \cite{MumfordI}.   From eq. (\ref{148}) one can see that the first requirement is satisfied if
\begin{equation}
F^{(yy)}=0.
\end{equation}
In appendix \ref{wave-function} it is shown that also the second condition is fulfilled. Summarizing, the wave-function as written in eqs. (\ref{ansatz1}) and (\ref{sol}) is the solution of the Laplace-Beltrami operator with the boundary conditions imposed by the torus geometry only if the components $xx$ and $yy$ of the magnetic field are vanishing \cite{0906.3033}.

Finally, in order to satisfy the first line of eq. (\ref{IIbc}), one has also to impose:
\begin{eqnarray}
 I^{t}  \vec{j}  = \vec{m} \in \mathbb{Z}^{d} \label{aparameter}
\end{eqnarray}
with $\vec{j}\in \mathbb{Z}^d$.
By introducing  the unitary vectors ${e}_i^t=(\overbrace{0,\dots,0, 1}^{i-times},0,\dots)^t$  in $\mathbb{Z}^d$ and writing $\vec{m}~ = ~m^{i} \vec{e}_{i}$, one can define the lattice \cite{0904.0910}
\begin{eqnarray}
\vec{j} = m^i \vec{j}_i~~;~~\vec{j}_i = I^{-t} \vec{e}_i \,\, .  \nonumber
\end{eqnarray}
Different values of $\vec{j}$ give different wave functions associated to chiral states having the same mass. Therefore  $\vec{j}$ labels the mass degeneracy of the ground states.
The Riemann Theta function is invariant under the translation $\vec{j}\rightarrow\vec{j}+\vec{e}_i$ and the inequivalent values of $\vec{m}$, namely the values of $\vec{m}$ which lead to different wave-functions,  are those inside the cell determined by the vectors $\vec{e}_i$. The determinant of the matrix $I$ which connects the two sets of vectors $\vec{j}_i$ and $\vec{e}_i$ provides the quantitative measure of the inequivalent $\vec{m}$'s and therefore gives the degeneracy of the ground state.

The full wave-function of the ground state, in the real coordinates system and in the case $F^{(xx)}=F^{(yy)}=0$, is
\begin{eqnarray}
\! \! \!  \! \! \! \phi_0\equiv\phi_{\vec{j}}^{ab}= {\cal C}_{ab}\,e^{i\vec{y}^t\,\frac{I_{ab}}{(2\pi R)^2} \, \vec{x}+i \vec{y}^t \Omega_{ab}^t\,\frac{  I_{ab}^t }{(2\pi R)^2}\vec{y}}\sum_{\vec{n}\in \mathbb{Z}^{d}}
e^{ i \pi (\vec{n}+\vec{j})^t  I_{ab} \Omega_{ab}  (\vec{n}+\vec{j})+  2 i \pi (\vec{n}+\vec{j})^t  I_{ab} (\frac{\vec{x}+\Omega_{ab} \, \vec{y}}{2\pi R})}\label{83}
\end{eqnarray}
with ${\cal C}_{ab}$ being an arbitrary overall constant. The choice of this constant affects the normalization of the kinetic terms of the scalars. In appendix \ref{wave-function} it is shown that canonically normalized kinetic terms are obtained by taking
\begin{eqnarray}
{\cal C}_{ab}= \sqrt{2g}V_{T^{2d}}^{-1/2} \left[ {\rm det}( I_{ab}{\rm Im}\Omega_{ab})\right]^{1/4} \nonumber
\end{eqnarray}
being $V_{T^{2d}}$  the torus volume.
It is also interesting to observe that
\begin{eqnarray}
\phi_{\vec{j}}^{ba}(\Omega_{ba})=(\phi_{\vec{j}}^{ab}(\Omega_{ab}))^*    \nonumber
\end{eqnarray}
which follows from the identities $I_{ab}=-I_{ba}$ and $\Omega_{ba}=(\bar{\Omega}_{ab})^*$. This is a consequence of the definition of $\bar{\Omega}$, written  after  eq. (\ref{defcl}).

The wave-function (\ref{83}) can be easily compared with eq. (5.39) of ref. \cite{0904.0910} where the negative chirality wave-function on the torus $T^4$ with trivial complex structure is given. The two expressions coincide if the generalized complex structure here introduced is identified with the modular matrix $i\hat{\Omega}$ defined in that reference. Such identification will become more transparent in the next section where  it will be explicitly  shown that in a torus $T^4$ with complex structure $U=i\mathbb{I}$ the generalized complex structure reduces to the modular matrix written, for example, in eq. (5.43) of ref. \cite{0904.0910}.

Eq. (\ref{83}) can be also compared with the wave-function of the chiral scalars defined on the factorized torus $(T^2)^d$ and given for example in ref. \cite{0701292}. On the factorized torus, the background gauge field is  already a block diagonal matrix  and the signs of its eigenvalues coincide with the ones of the first Chern classes. In this background, the generalized complex structure is the following diagonal matrix:
\begin{eqnarray}
\Omega\equiv {\rm diag}\left(\dots , \left(\frac{1+\mbox{sign} \, \lambda_r}{2}\right) U_r+  \left(\frac{1-\mbox{sign}\,  \lambda_r}{2}\right)\bar{U}_r, \dots \right) \,\, . \label{gcs}
\end{eqnarray}
The $U_r$s are the complex structures of the single component $T^2$ of the factorized torus $(T^2)^d$.
The corresponding wave-function will be the factorized product of $d$ functions, each of them depending on
the holomorphic $U_r$ or anti-holomorphic  $\bar{U}_r$ variables, according to the sign of  $\lambda_r$. Such wave-function, whose  holomorphicity properties are related to the signs of the first Chern classes, coincides with the one introduced in ref. \cite{0810.5509}.

Summarizing, the solution of the Laplace-Beltrami equation on a torus $T^{2d}$ with arbitrary complex structure has been found. This solution provides the internal
wave-function of the chiral scalars corresponding to the open strings ending on two stacks of magnetized branes. The difference in their magnetization is a matrix having  non-vanishing elements only along the non-diagonal blocks. Under this assumption  and  after having chosen a suitable ansatz for the internal wave function, the boundary conditions dictated by the magnetized torus geometry reduces to the quasi periodicity conditions  satisfied by the Riemann Theta function. Let us notice that such assumption, in the case of only one dy-charged sector  is not restrictive because a generic antisymmetric matrix, associated to the unique  difference $F^{ab}$ of the magnetic fields on the two piles $a$ and $b$, can be always recast in  a matrix having vanishing elements along the diagonal blocks through an orthogonal rotation. But in the case of more dy-charged sectors, it is not possible to put {\em all}  the differences of the magnetic fields simultaneously in that particular form unless  extra conditions are imposed. That is why, being interested in the computation of the Yukawa couplings, where three wave-functions are involved, we have imposed  the vanishing of the diagonal blocks of the magnetic fields.   This is reminiscent of what happens in the stringy calculus of the Yukawa couplings on $T^{6}$ \cite{0709.1805} that is, in fact,  performed under the assumption that all the monodromy matrices associated to the different dy-charged sectors are commuting.

In principle, it should be possible  to find the solution of the Laplace equation also in the most general case of a background magnetic field described by a matrix with all  the entries different from zero. From the analysis here performed, it seems that this solution should not be the Riemann Theta function even if a final answer to this question needs a more exhaustive study.

\subsection{The Dirac equation}
\label{Dirac equation}

The ten-dimensional fermion kinetic term  of the ${\cal N}=1$ SYM theory supported by the world-volume of a stack of $N$ D9-branes is
\begin{eqnarray}
S_2^{\Psi}= \frac{i}{2g^2} \int d^4x \sqrt{G_4}\int d^6X^N \bar{\Psi}^{ba}\left[(\Gamma^{\mu}D_\mu+\Gamma^M\tilde{D}_M)\Psi\right]^{ab} \label{actD9}
\end{eqnarray}
where, in view of a subsequent dimensional reduction, the internal compact directions have been separated from the four-dimensional ones. Compactification of the action (\ref{actD9}) is obtained by first decomposing the Dirac matrices in the factorized product of the $D=4$ and $D=6$ representations of the Clifford algebra, as shown in eq. (\ref{163}),
and then by expanding the ten-dimensional fermion fields in terms of the wave-functions which are solutions of the internal Dirac equation, with the boundary conditions implied by the compact geometry \cite{0404229}
\begin{eqnarray}
\Psi^{ab}= \sum_n\psi_n^{ab}\otimes\eta_n^{ab}~~;~~i \gamma^M_{(6)}\tilde{D}_M\eta_n^{ab}=m_n\eta_n^{ab} \,\, .\label{Dirac}
\end{eqnarray}
In analogy with the dimensional reduction of the bosonic kinetic terms, the eigenfunction problem of the Dirac equation is solved in the complex frame ${\cal Z}$ where both the metric and the magnetic background are diagonal in the non-vanishing off-diagonal blocks: see eqs. (\ref{cmpxmtr}) and (\ref{89}).
In this complex frame, the Clifford algebra becomes:
\begin{eqnarray}
\left\{\gamma^{{\cal Z}^r},\, \gamma^{\bar{\cal Z}^s}\right\}= 4 \delta^{rs}   \nonumber
\end{eqnarray}
while all the remaining anti-commutators vanish. This algebra is the usual one of fermion creation and annihilation operators and the gamma-matrices can be identified with such operators. According to the identifications (\ref{osc0}), also the covariant derivatives satisfy, apart from a  factor, the algebra of the bosonic creation and annihilation operators.  Then,  the massless state living in the kernel of the Dirac equation
is obtained by defining a factorized  vacuum  $
\eta_0(\vec{\cal Z}\, , \,{\vec{\bar{\cal Z}}})~=~ u_0~\otimes~\phi_0(\vec{\cal Z}\, ,\,{\vec{\bar{\cal Z}}})$.
Here,  $u_0$ is a constant six-dimensional spinor  and $\phi_0$ is a function of the internal coordinates, both vanishing under the action respectively  of  all the  fermionic and bosonic annihilation operators
\begin{eqnarray}
D_{r}^{({\bar{\cal Z}})}\phi_0({\vec{\cal Z}}, {\vec{\bar{\cal Z}}} )=0~~;~~\gamma^{{\cal Z}^r}_{(6)}u_0=0~~;~~{\rm for}~ \lambda_r>0 \nonumber \\
D_{r}^{{({\cal Z})}}\phi_0({\vec{\cal Z}}, {\vec{\bar{\cal Z}}} )=0~~;~~\gamma^{\bar{\cal Z}^r}_{(6)}u_0=0~~;~~{\rm for}~ \lambda_r<0 \label{soldirac}
\end{eqnarray}
together with the boundary conditions given in eq. (\ref{bctorus}).
Eqs. (\ref{soldirac}) show that for positive $\lambda_r$ the holomorphic gamma-matrices   $\gamma^{{\cal Z}^r}_{(6)}$ have  to be identified with the  annihilation operators, while the anti-holomorphic Dirac matrices play the same role for opposite signs. This sign-dependent identification can be avoided  by rewriting the gamma-matrices in the $Z$-frame where the coordinates are seen as holomorphic or anti-holomorphic according to the sign of  $\lambda_r$ (see eq. (\ref{clcoo})). In this frame  one always has  that $\gamma^{{Z}^r}_{(6)}$ and $\gamma^{\bar{ Z}^r}_{(6)}$
are identified respectively with the annihilation and creation operators.

The solution of  eq. (\ref{soldirac}) is then obtained  by assuming $\phi_0$ to be  the wave-function in eq. (\ref{83}) and by defining  $u_0= \gamma^{{\cal Z}^r}\chi_0$, for positive eigenvalues $\lambda_r$ and   $u_0= \gamma^{\bar{\cal Z}^r}\chi_0$ for negative eigenvalues,  being $\chi_0$ an arbitrary eight-component spinor. In the $Z$-coordinates we have:
\begin{eqnarray}
\eta_0(\vec{Z},\,\vec{\bar Z})=\prod_{r=1}^3 \gamma^{{Z}^r}\chi_0\otimes \phi_{\vec{j}}(\vec{Z},\,\vec{\bar Z})\label{vacccum} .
\end{eqnarray}
This spinor is chiral. In appendix \ref{torus} it is explicitly shown that it has positive chirality when, in the ${\cal Z}$ coordinates, an odd number of Dirac matrices with anti-holomorphic indices appear in its definition, while it has negative chirality in the other case.  The signs of the $\lambda_r$s also  determine which combinations of gamma-matrices appear in the definition of the massless states and for this reason one can verify  that  the six-dimensional chirality is given by  the product of such signs. Finally, those signs also impose the wave-function and its complex conjugate to have opposite chirality. This is  also true  in the case of the $T^2$-torus \cite{0404229}.

The whole spectrum of the Kaluza-Klein fermions is obtained, following the standard procedure, by squaring  eq. (\ref{Dirac}):
\begin{eqnarray}
-\left(\gamma^{{\cal Z}^I}_{(6)} D_{I}^{\cal{Z}}\gamma^{{\cal Z}^J}_{(6)} D_{J}^{\cal{Z}}\right)\eta_n & = &
\sum_{r=1}^3\left( |\lambda_r|(2N_r+1) -\frac{1}{4} \left[ \gamma^{{ Z}^r}_{(6)},\,\gamma^{\bar{ Z}^r}_{(6)}\right]|\lambda_r|\right) \eta_n \nonumber \\
&= & (2 \pi R)^{2} m_n^2\eta_n \, , \label{lapl}
\end{eqnarray}
where the bosonic number operator, defined in the previous sections, has been introduced  and  the expression of the background gauge field given in eq.~(\ref{29}) has been used.

The vacuum state shown in eq. (\ref{vacccum}) satisfies the previous equation with $m=0$ and, applying on it an arbitrary number of bosonic oscillators
\begin{eqnarray}
(a_1^\dagger)^{N_1}\,(a_2^\dagger)^{N_2}\,(a_3^\dagger)^{N_3}  \prod_{r=1}^3\gamma^{{Z}^r}\chi_0\otimes \phi_{\vec{j}}(\vec{Z},\,\vec{\bar Z})\,\, ,   \nonumber
\end{eqnarray}
a set of Kaluza-Klein states are generated with masses
\begin{eqnarray}
m^2 =\frac{2}{(2\pi R)^2} \sum_{r=1}^3|\lambda_r| N_r.   \nonumber
\end{eqnarray}
The next level in the fermion Fock space, satisfying eq. (\ref{lapl}),  is  obtained by applying  a fermion creation operator
 on the massless state\begin{eqnarray}
\gamma^{\bar{Z}^k}  \prod_{r=1}^3\gamma^{{Z}^r}\chi_0\otimes \phi_{\vec{j}}(\vec{Z},\,\vec{\bar Z})~~;~~k=1,2,3.\label{first}
\end{eqnarray}
These are three states with  mass $m^2= 2 |\lambda_k|/(2\pi R)^2$. Their chirality  can be determined first by defining   $\chi= \gamma^{Z^k}\chi_0$ as a general six-dimensional spinor and then counting, in the ${\cal Z}$ coordinates,  the number of holomorphic Dirac matrices acting on it. This number differs by one from the number of the same matrices acting on the vacuum.  The relation between chirality and the number of holomorphic Dirac matrices makes one conclude that the states (\ref{vacccum}) and (\ref{first})   have necessarily opposite chirality. Finally, by acting  on this level with an arbitrary number  of  bosonic creation operators:
\begin{eqnarray}
(a_1^\dagger)^{N_1}\,(a_2^\dagger)^{N_2}\,(a_3^\dagger)^{N_3} \gamma^{\bar{Z}^k}  \prod_{r=1}^3\gamma^{{Z}^r}\chi_0\otimes \phi_{\vec{j}}(\vec{Z},\,\vec{\bar Z})~~~~k=1,2,3.  \nonumber
\end{eqnarray}
a tower of KK states is generated with  masses given by:
\begin{eqnarray}
m^2_k=\frac{1}{(2\pi R)^2} \sum_{r=1}^3|\lambda_r|(2N_r)+2\frac{|\lambda_k|}{(2\pi R)^2}~~~~~k=1,2,3 \,\,\,  .   \nonumber
\end{eqnarray}
Other  KK towers  are obtained by acting on the vacuum with two or three fermion creation oscillators and an arbitrary number of bosonic oscillators
\begin{eqnarray}
(a_1^\dagger)^{N_1}\,(a_2^\dagger)^{N_2}\,(a_3^\dagger)^{N_3} \gamma^{\bar{Z}^k} \gamma^{\bar{Z}^l}  \eta_0~~;~~(a_1^\dagger)^{N_1}\,(a_2^\dagger)^{N_2}\,(a_3^\dagger)^{N_3} \prod_{k=1}^3\gamma^{\bar{Z}^k}\eta_0   \nonumber
\end{eqnarray}
with $k,l=1,2,3$. These  are three and one tower of massive states having respectively the same and opposite chirality of the vacuum. Their mass spectrum is given by:
\begin{eqnarray}
m^2_{k,l}=\frac{2}{(2\pi R)^2} \sum_{r=1}^3|\lambda_r| N_r+2\frac{|\lambda_k|+|\lambda_l|}{(2\pi R)^2}~~;~~m^2=\frac{2}{(2\pi R)^2} \sum_{r=1}^3|\lambda_r| (N_r+1) \,\, .  \nonumber
\end{eqnarray}
All the mass formulas can be collected in a more concise relation by introducing the fermion number operator $N^f_r=0,1$ and by writing
 \begin{eqnarray}
 m^2_n=\frac{2}{(2\pi R)^2} \sum_{r=1}^3|\lambda_r| (N_r+N_r^f) \,\, .\label{massspectrum}
\end{eqnarray}
The masses of the KK states are parameterized by the bosonic and fermionic occupation numbers $(N_r,\,N_r^f)$.
States having occupation numbers $(N_r,\,N_r^f=0)$ and  $(N_r-1,\,N_r^f=1)$ have the same mass but opposite chirality and  they are the two components of a
four-dimensional  massive Dirac spinor.

The mass spectrum (\ref{massspectrum}), as explicitly shown in appendix \ref{string}, coincides with the zero-slope limit of the string mass formula in the R-sector and for dy-charged open strings.

All these states, solutions of the Dirac equation,  are  written in the ${\cal Z}$-frame
which has been introduced in order to diagonalize the off-diagonal blocks of the magnetic field and so it is necessarily dependent on its background values.
In configurations  involving several piles of magnetized branes  there are a lot of such frames, each of them associated to the sets of  dy-charged open strings stretched between different stacks of magnetized branes and therefore there are many wave-functions depending on different local system of coordinates. The calculation of the effective actions demands the evaluation of overlap integrals among three or more of these functions. One has to re-express such states in terms of quantities defined in a unique system of coordinates as the ${\cal W}^M$ one.
In this frame one can write:
\begin{eqnarray}
\eta_0(\vec{w},\,\vec{\bar w})=\prod_{r=1}^3 \left( C^{r}_{s} \frac{ (1+\mbox{sign} \lambda_{r})}{2} \gamma^{w^{s}} + \bar{C}^{r}_{s}
\frac{ (1- \mbox{sign} \lambda_{r})}{2} \gamma^{\bar{w}^{s}}
\right) \chi_0\otimes \phi_{\vec{j}}(\vec{w},\,\vec{\bar w})\label{vaccum}
\end{eqnarray}
where $C_{r}^{s}, \bar{C}^{r}_{s}$ are the inverse matrices of the ones  defined in eq. (\ref{eqei}) and $\phi_{\vec{j}}$  is a scalar function of the coordinates.

\subsection{Example: $T^4$ with complex structure $U=i\mathbb{I}_{4}$}

In order to compare the results here obtained with the ones in
literature, it is useful to specify the wave-function,
the equation of motions and the background gauge field for the torus  $T^4$ with the trivial complex structure \cite{0904.0910}.

The torus is defined by introducing, in $\mathbb{R}^4$, the identifications
\begin{eqnarray}
x^i\equiv x^i+2\pi R~~;~~y^i\equiv y^i+2\pi R~~;~~i=1,2  \nonumber
\end{eqnarray}
with the metric given by:
\begin{eqnarray}
ds^2=\delta_{ij}dx^i\, dx^j  + \delta_{ij}dy^i\, dy^j \,\, .   \nonumber
\end{eqnarray}
Complex variables are introduced in the standard way:
\begin{eqnarray}
z^i=\frac{x^i+ iy^i}{2\pi R} \,\, .    \nonumber
\end{eqnarray}
These define a complex torus with  complex structure $U=i\mathbb{I}_4$. The non-vanishing block of the difference $F^{ab}$ between the background magnetic fields activated on the world-volume of the two piles $a$ and $b$, for any sector $ab$, is:
\begin{eqnarray}
F^{(xy)} =\left( \begin{array}{cc}
                     f_{11}&f_{12}\\
                     f_{12}&f_{22}\end{array}\right) \,\, .   \nonumber
\end{eqnarray}
This matrix is symmetric  because of the  eq. (\ref{15}) which implies $\tilde{F}= F^{(xy)}\,\,  {\rm Im }U^{-1}=F^{(xy)}$ and of the condition $\tilde{F}^\dagger=\tilde{F}$ which becomes $\tilde{F}^t=\tilde{F}$ on a real matrix.

The eigenvalues of this matrix are real and given by
\begin{eqnarray}
\lambda_{\pm} = \frac{f_{22}+f_{11}}{2}\pm \frac{1}{2} \sqrt{(f_{22}-f_{11})^2 + 4f_{12}^2}   \nonumber
\end{eqnarray}
and the corresponding orthonormal eigenvectors  are:
\begin{eqnarray}
v_+ =\frac{1}{\sqrt{a^2+1}} \left(\begin{array}{c}
                -a\\1\end{array}\right) ~~;~~  v_- =\frac{1}{\sqrt{a^2+1}} \left( \begin{array}{c}
                                                               1\\ a\end{array}\right)   \nonumber
\end{eqnarray}
with $a= \frac{f_{22}-f_{11}}{2f_{12}}- \frac{1}{2f_{12}} \sqrt{(f_{22}-f_{11})^2 + 4f_{12}^2}$ which is a real quantity because the argument of the square root  is positive definite.
These eigenvectors, according to eq.~(\ref{eqei}), are collected in the matrix
\begin{eqnarray}
C^{-1}=C^t=C=\frac{1}{\sqrt{a^2+1}} \left( \begin{array}{cc}
                                                -a&1\\
                                                1&a\end{array}\right)\label{122}
\end{eqnarray}
and the non-vanishing element of the gauge field becomes:
\begin{eqnarray}
\tilde{F} & = &  C^t\frac{\mathbb{I}_\lambda}{(2\pi R)^2} \bar{C} \nonumber \\
& = &
\frac{\lambda_-}{ (2\pi R)^2(a^2+1)}\left(\begin{array}{cc}
                                                                 1&a\\
                                                                 a & a^2\end{array}\right)
+\frac{\lambda_+}{ (2\pi R)^2(a^2+1)}\left(\begin{array}{cc}
                                                                 a^2&-a\\
                                                                 -a&1\end{array}\right) \, .\label{95}
\end{eqnarray}
Eq. (\ref{95})  has the same structure as eq. (5.7) in ref. \cite{0904.0910}.  In the following, just to fix the notation, we assume that $\lambda_+$ is positive while $\lambda_-$ is negative. However, according to the definition of the matrix $C^{(\lambda)}$ given in eq.  (\ref{defcl}), being $C$ real, we have $C^{(\lambda)}=C$ and
\begin{eqnarray}
\Omega={C^{(\lambda)}}^{-1}\tilde{C}^{(\lambda)}  = & \frac{i}{a^2+1} \left( \begin{array}{cc}
                                                                              -a&1\\
                                                                              1&a\end{array}\right)\left(\begin{array}{cc}
                                                                                    -a&1\\
                                                                                    -1&-a\end{array}\right)
\nonumber \\ = & - \frac{i}{a^2+1}\left( \begin{array}{cc}
                 1-a^2& 2a\\
                 2a &a^2-1\end{array}\right) \,\,. \label{124}
\end{eqnarray}
Again, this equation is formally identical to eq. (5.43) of ref. \cite{0904.0910}.  By using it together with  the definition of $C^{(\lambda)}$ and eq. (\ref{148}), it is simple to compute $F^{(xy)}$ and to check that it is equal to eq. (\ref{95}).

The solution of the Laplace-Beltrami equation, according to the general analysis made in the previous section, is equivalent to solve:
\begin{eqnarray}
D_{1}^{(\bar{\cal Z})}\phi_0=0~~;~~D_{2}^{({\cal Z})}\phi_0=0  \label{ZZ}
\end{eqnarray}
with the boundary conditions given by eq. (\ref{bctorus}). When written in terms of the flat  coordinates $w, \bar{w}$ introduced in eq. (\ref{19}), eqs. (\ref{ZZ}) become:
\begin{eqnarray}
&&\left(D_{1}^{(\bar{w})}\,(C^{-1})^1_{~1}+ D_{2}^{(\bar{w})}\, (C^{-1})^2_{~1}\right)\phi_0= (-a\,D_{1}^{(\bar{w})}+D_{2}^{(\bar{w})})\phi_0=0 \,\, , \nonumber\\
&& \left( D_{1}^{({w})}\,(C^{-1})^1_{~2}+D_{2}^{(w)}\, (C^{-1})^2_{~2}\right)\phi_0= (D_{1}^{(w)}+aD_{2}^{(w)})\phi_0=0\label{74} .
\end{eqnarray}
These  are formally the same as eqs. (5.4) and (5.5) of ref. \cite{0904.0910}.

The Dirac equation, instead, is solved by rewriting $\eta_{0}$ as follows:
\begin{eqnarray}
\eta_0({x}^n,{y}^n)=u_0\phi_0({x}^n,{y}^n)    \nonumber
\end{eqnarray}
where $\phi_0({x}^n, {y}^n)$ satisfies eqs. (\ref{74}) and  $u_0$ is a constant spinor annihilated by the fermion destruction operators:
\begin{eqnarray}
\gamma^{{\cal Z}^1}u_0=0~~;~~\gamma^{\bar{\cal Z}^2}u_0=0 \,\,\, .   \nonumber
\end{eqnarray}
The solution of these equations is:
\begin{eqnarray}
u_0= \gamma^{{\cal Z }^1}\gamma^{\bar{\cal Z }^2} \lambda\,\,\, ,  \nonumber
\end{eqnarray}
$\lambda$ being an arbitrary spinor. By rewriting it in the system of coordinates $w$ one gets:
\begin{eqnarray}
u_0&=& \left( {C}^1_{~1}\gamma_{(4)}^{{w}^1} + {C}^1_{~2}\gamma_{(4)}^{{w}^2}\right)
\left( \bar{C}^2_{~1}\gamma_{(4)}^{ \bar{w}^1} + \bar{C}^2_{~2}\gamma_{(4)}^{\bar{w}^2}\right)\lambda\nonumber\\
&=& \frac{1}{a^2+1} \left( \begin{array}{cccc}
                   0&0     &0    &0\\
                   0&4a &-4 &0\\
                   0&4a^2    & -4a &0\\
                   0&0     &0    &0\end{array}\right)\left(\begin{array}{c}
                                                              \lambda_1 \\
                                                              \lambda_2\\
                                                              \lambda_3\\
                                                              \lambda_4\end{array}\right)=
\left(\begin{array}{c}
                                                              0 \\
                                                              4a\lambda_2-4\lambda_3\\
                                                              4a^2\lambda_2-4a\lambda_3\\
                                                              0\end{array}\right)\label{spinsol}
\end{eqnarray}
where the following representation of the Dirac matrices has been used:
\begin{eqnarray}
\gamma_{(4)}^{w^1}= \gamma^{w}\otimes \sigma^3~~;~~\gamma_{(4)}^{\bar{w}^1}=\gamma^{\bar{w}}\otimes \sigma^3  \nonumber\\
\gamma_{(4)}^{w^2}= \mathbb{I}\otimes \gamma^{w}~~;~~\gamma_{(4)}^{\bar{w}^2}=\mathbb{I}\otimes \gamma^{\bar{w}} \,\, .  \nonumber
\end{eqnarray}
The matrices $\gamma^{w,\bar{w}}$  are given in eq. (\ref{curvega}) and $\sigma^3$ is one of the Pauli matrices. The internal chirality, instead, turns out to be $\gamma^5=\sigma^3\otimes\sigma^3$. The spinor (\ref{spinsol}) has negative chirality and the ratio of its  two non-vanishing components  is $1/a$. By comparing it  with the same ratio of the two components of the spinor (5.18) of  ref. \cite{0904.0910}, we get completely agreement with $q=1/a$. This value makes $\Omega$, given in eq. (\ref{124}),  coincident with $\Omega$ introduced in ref. \cite{0904.0910} (see eq. (5.43)). Finally, by using it again in eq. (\ref{95}) and identifying $\lambda_+ \equiv (2\pi R)^2(1+q^2)\hat{N}_{1\bar{1}}$ and $\lambda_-\equiv (2\pi R)^2(1+q^2)\hat{N}_{2\bar{2}}$, one can see that this equation becomes identical to eq. (5.7) in ref. \cite{0904.0910}.  The wave-function given in eq. (\ref{83}) can now be easily compared with the second line of eq. (5.39) of the aforementioned paper.

\section{Yukawa couplings  for magnetized branes compactified on $T^6$}
The general expression of the four-dimensional  Yukawa couplings, involving chiral dy-charged matter
has been obtained, in the framework of magnetized branes, in ref. \cite{0404229}. Here we give the result:
\begin{eqnarray}
S_{3}^{\Phi} &=&  \frac{1}{2g^2}
\int d^4 x \sqrt{G_{4}}  \int d^6 X^N \sqrt{G_{6}}
\bar{\psi}^{ca}_0({x}^\mu)\,\gamma^5_{(4)} \left[ \varphi_{i,\,0}^{ab}({x}^\mu)\,
\psi_0^{bc}({x}^\mu)\otimes (\eta^{ac}_0)^\dag(x^n,{y}^n)\right.  \nonumber \\
&\times&
\gamma^i_{(6)} \phi^{ab}_{0}(x^n,{y}^n) \eta_0^{bc}(x^n,{y}^n)  - \varphi_{i,\,m}^{bc}({x}^\mu)\,
\psi_0^{ab }({x}^\mu)\otimes (\eta^{ac}_0)^\dag(x^n,{y}^n)\nonumber\\
&\times&\left.\gamma^i_{(6)} \phi^{bc}_{0}(x^n,{y}^n) \eta_0^{ab}(x^n,{y}^n)
\right]\nonumber\\
\label{S31}
\end{eqnarray}
where $\psi_0$ and $\varphi_{0}$ are respectively the lightest fermionic and bosonic  excitations of the   Kaluza-Klein towers. These are massless in the case of the fermions while the lightest boson becomes massless only if a supersymmetry condition is imposed.
In the following, in order to fix notations, we choose $\lambda_1^{ab}$ to be positive and we  impose the supersymmetry  condition:
\begin{eqnarray}
|\lambda_1^{ab}|=|\lambda_2^{ab}|+|\lambda_3^{ab}|   \nonumber
\end{eqnarray}
where the $\lambda$s are the eigenvalues of the difference of the background magnetic fields living on the branes labeled with the $a,b$ indices.  The massless scalar with this choice of the magnetic field  is $\phi_{{\cal Z}^1}$, while with the opposite choice of $\lambda_1$ is $\phi_{\bar{\cal Z}^1}$. In the chosen notations, only the first term in eq. (\ref{S31}) contributes to the Yukawa coupling for massless particles and one is left with the expression
\begin{eqnarray}
(S_{3}^{\Phi})^{(1)}= \int d^4 x \sqrt{G_{4}}  \bar{\psi}^{ca}_0\,\gamma^5_{(4)}
\varphi_{{\cal Z}^1,0}^{ab}\, \psi_{0}^{bc} Y^{s}
\label{YY}
\end{eqnarray}
with the Yukawa coupling constants, in the string frame, given by
\begin{eqnarray}
Y^{s} &=& \frac{1}{2g^2} \left[(u^{ac}_0)^\dag \gamma^{{\cal Z}^1_{ab}}_{(6)} u_0^{bc}\right]{\cal Y}^s\label{yukcoup}
\end{eqnarray}
being
\begin{eqnarray}
{\cal Y}^s&=&\int_{T^6}d^3x d^3 y \sqrt{G_6}(\phi^{ac}_0(\vec{Z}^{ac},\,\vec{\bar Z}^{ac}))^\dag
\phi^{ab}_{0}(\vec{Z}^{ab},\,\vec{\bar Z}^{ab}) \phi_0^{bc}(\vec{Z}^{bc},\,\vec{\bar Z}^{bc}) \,\, .
\label{YY2}
\end{eqnarray}
The spinors $u_0$ and $\psi_0$ are defined in eqs. (\ref{soldirac}) and (\ref{Dirac}) where, as explained in sect. [\ref{reduction and fluxes}], the magnetic dependence of the complex coordinates has been emphasized by labeling them with the indices specifying the stacks of branes where the corresponding open strings have their endpoints.

The spinor product in eq.  (\ref{yukcoup}) determines which fermions do not have vanishing couplings. In fact, such product is different from zero only if the two constant spinors have opposite chirality. This feature, shown to be true in the case of the factorized torus $(T^2)^3$ in ref. \cite{0810.5509}, can be proven by first observing that the complex Dirac matrices anticommute with the chirality operator
\[
\left\{\gamma_{(6)}^7\,,\gamma^{{\cal Z}^1_{ab}}_{(6)}\right\}=(C_{ab})^1_{~r}\left\{\gamma_{(6)}^7\,,\gamma^{w^r}_{(6)}\right\}=0
\]
 and by computing the product
\begin{eqnarray}
(u^{ac}_0)^\dag \gamma^{{\cal Z}^1_{ab}}_{(6)}u_0^{bc}=(-1)^{\eta_{bc}}(u^{ac}_0)^\dag \gamma^{{\cal Z}^1_{ab}}_{(6)} \gamma_{(6)}^7u_0^{bc}
= -(-1)^{\eta_{bc}+\eta_{ac}}(u^{ac}_0)^\dag \gamma^{{\cal Z}^1_{ab}}_{(6)}u_0^{bc}  \nonumber
\end{eqnarray}
which is different from zero only if the two spinors have opposite chirality. Here, we have denoted by
$(-1)^\eta$ ($\eta=0,1$) the chirality of the constant spinor.

The first factor in eq. (\ref{yukcoup}) also determines the holomorphy of the wave-functions that give a non-zero Yukawa coupling.  With our choice of the massless scalar, the definition of fermion vacuum given in eq. (\ref{vaccum}) leads to a non-vanishing coupling  only by taking $\eta_0^{bc}$ either as the product of three anti-holomorphic Dirac matrices acting on a general spinor (all the $\lambda_{r}^{bc}$'s are negative and $\Omega_{bc}=\bar{U}$) or as a mixed product of holomorphic and anti-holomorphic Dirac matrices acting again on a general spinor (the signs of the $\lambda_{r}^{bc}$'s are mixed).
In the latter case the matrix $\Omega_{bc}$ is neither equal to the complex structure of the torus nor to its complex conjugate.
Similar consideration can be given to the spinor $(u^{ac}_0)^\dag$ with the proviso that this spinor has to be taken with opposite chirality with respect the one of $u_0^{bc}$.
It is worth to observe that, even taking $\eta_0^{bc}$ completely anti-holomorphic and the scalar completely holomorphic according to our initial choice of $\lambda_1^{ab}>0$, one cannot use as   $(u^{ac}_0)^\dag$ the product of holomorphic or anti-holomorphic Dirac matrices, because their algebra would necessarily give a zero result.
Non-vanishing Yukawa couplings  have then  to involve an overlap of wave functions where at least one
of the three matrices $\Omega$  is different from the torus complex structure $U$ or from its complex conjugate $\bar{U}$.
For this reason, an  overlap integral among  wave functions depending on three arbitrary matrices $\Omega$ is now going to be evaluated.  Details of the calculation are given in appendix \ref{Yukawa calculus}. The result follows:
\begin{eqnarray}
{\cal Y}^{s} & = &  \int d^3\tilde{x} d^3\tilde{y}\sqrt{G_6}
{\phi_{\vec{j}_1}^{ca}(\Omega_{ca})}^*
\phi_{\vec{j}_1}^{ab}(\Omega_{ab})\, \phi_{\vec{j}_2}^{bc}(\Omega_{bc})
={\cal N}_{ab}{\cal N}_{bc}{\cal N}_{ca}\sqrt{G_6}\nonumber\\
&\times&\, {\cal D}\,\left[{\rm det}(-i(I_{ca}\Omega_{ca}+I_{ab}\Omega_{ab}+I_{bc}\Omega_{bc}))\right]^{-1/2}\nonumber\\
&\times&
\sum_{\substack{\vec{p}\in\mathbb Z^{3}_{{\rm det}[I_{bc}] I_{bc}^{-1} }\\ \vec{\tilde p}\in  \mathbb Z^{3}_{{\rm det}[I_{ab}] I_{ab}^{-1} }  }}
\Theta\left[\substack{ \alpha^{-t}I_{bc}^t (\vec{j}_3-\vec{j}_2)+\frac{I_{bc}^t}{{\rm det} I_{bc}^t}\vec{p}+\frac{I_{ab}^t}{{\rm det}I_{bc}}\vec{\tilde{p}}
\\0}\right](0| \Pi)\label{yukt6}
\end{eqnarray}
with
\begin{eqnarray*}
\Pi&=&\alpha\left((\Omega_{ab}I_{ab}^{-t}+\Omega_{bc}I_{bc}^{-t})-(\Omega_{ab}-\Omega_{bc})
(I_{ca} \Omega_{ca}+I_{ab} \Omega_{ab}+I_{bc}\Omega_{bc})^{-1}(\Omega_{ab}-\Omega_{bc})^t\right)
\alpha^{t}\\
{\cal D}&=&\frac{1}{ \chi}\sum_{\vec{m}\in\tilde{\mathbb{Z}}_{(I_{ab}^{-1}+I_{bc}^{-1})\alpha}}
\delta_{(\vec{j}_1^t\,I_{ab}+\vec{j}_2^t\,I_{bc}+\vec{m}^t\,I_{ab})(I_{ab}+I_{bc})^{-1}; \vec{j}_3^t}
\end{eqnarray*}
being $\chi$  defined in eq. (\ref{chi}) and $\alpha=\mbox{det}\, [{I}_{ab}{I}_{bc}]\mathbb{I}$ \cite{0904.0910} . The matrix $\alpha$ has been introduced in eq. (\ref{mdef})
of appendix \ref{Yukawa calculus} to make the product $\alpha({I}_{ab}^{-1}+{I}_{bc}^{-1})$ an integer matrix. This has been essential in getting an identity between the product of two Riemann Theta functions. As in the case of  the calculus of the Yukawa couplings on the factorized torus, it is the main ingredient to compute the overlap integral of three wave-functions giving the Yukawa couplings on the general torus.

Eq. (\ref{yukt6}) simplifies when all the differences of magnetic fields living on the various stacks of magnetized branes are independent but commuting.  An analogous string calculus of the Yukawa coupling has been performed in ref. \cite{0709.1805}.  In such a configuration, it is convenient to introduce the matrix having integer entries  $P~=~\alpha I_{ab}^{-1}I_{bc}^{-1}$ and to observe that:
\begin{eqnarray}
\hspace{-1cm} &&\sum_{\substack{\vec{p}\in\mathbb Z^{3}_{{\rm det}[I_{bc}] I_{bc}^{-1} }\\ \vec{\tilde p}\in  \mathbb Z^{3}_{{\rm det}[I_{ab}] I_{ab}^{-1} }  }}
\Theta\left[\substack{ \frac{I_{bc}^t}{{\rm det} I_{ab}I_{bc}}(\vec{j}_3-\vec{j}_2)+\frac{I_{bc}^t}{{\rm det} I_{bc}^t}\vec{p}+\frac{I_{ab}^t}{{\rm det}I_{bc}}\vec{\tilde p}
\\0}\right](0| P\,\tilde{\Pi}\,P^t)
=\Theta\left[\substack{ I_{ab}^{-t}(\vec{j}_3-\vec{j}_2)
\\0}\right](0| \tilde{\Pi})\label{yukcom}
\end{eqnarray}
with
\begin{eqnarray}
\tilde{\Pi} & = & I_{ab}I_{bc} \left((\Omega_{ab}I_{ab}^{-t}+\Omega_{bc}I_{bc}^{-t}) \right .\nonumber \\
& & - \left. (\Omega_{ab}-\Omega_{bc})
(I_{ca} \Omega_{ca}+I_{ab} \Omega_{ab}+I_{bc}\Omega_{bc})^{-1}(\Omega_{ab}-\Omega_{bc})^t\right)
I_{ab}^tI_{bc}^t \,\, .  \nonumber
\end{eqnarray}
More details on the identity written in eq. (\ref{yukcom}) are given at the end of appendix \ref{Yukawa calculus}.

Eq. (\ref{yukcom}) could have been directly obtained by performing a different choice of the matrix  $\alpha$.  When all the first Chern classes commute, the quantity $\alpha(I_{ab}^{-1}+I_{bc}^{-1})$ can be made an integer matrix by choosing $\alpha=I_{ab} I_{bc}$. With this choice, as it will be explicitly shown in appendix \ref{Yukawa calculus},  there is no need to introduce the vectors $\vec{p}$ and $\vec{\tilde{p}}$.
Furthermore, the quantity $\Pi \equiv \tilde{\Pi}$ and the characteristic of the Theta function comes out already in the form $I_{ab}^{-t}(\vec{j}_3-\vec{j}_2)$.

\subsection{Example: Yukawa couplings on factorized torus.}

The expression of the Yukawa couplings derived in the main section describes also the factorized torus $T^6~=~T_2^3$ where the background gauge field $F$ is
\begin{eqnarray}
F=F_{12}dx^1\wedge dy^1+F_{34} dx^2\wedge dy^2+F_{56}dx^3 \wedge dy^3  \nonumber
\end{eqnarray}
while the metric is
\begin{eqnarray}
\frac{ds^2}{(2\pi R)^2}=\sum_{r=1}^3\frac{{\cal T}_2^r}{U_2^r}\frac{|dx^{2r-1}+U^rdy^{2r}|^2}{(2\pi R)^2} \,\, .  \nonumber
\end{eqnarray}
Here, $U_{2}^r$ and ${\cal T}_{2}^r$ are respectively the imaginary parts of the complex and the K\"ahler structure of the two tori  $T_2^r$
in $T^6$ ($r=1,2,3$).

The complex frame $z$, where the metric and the magnetic field strength are trivial,  as in eqs. (\ref{metr}) and  (\ref{89}), is introduced through  the definitions
\begin{eqnarray}
z^r= \sqrt{\frac{{\cal T}_2^r}{U_2^r}}\frac{x^{r}+U^r\,y^{r}}{2\pi R}\equiv \sqrt{\frac{{\cal T}_2^r}{U_2^r}}w^r\equiv C^r_m w^m \,\, .   \nonumber
\end{eqnarray}
The real matrix $C$ is diagonal and coincides with  $C^{(\lambda)}$.
The gauge field in the complex frame is already diagonal in its non-vanishing blocks
\begin{eqnarray}
F= \sum_{r=1}^3\frac{U_2^r 2\pi I^{r}}{T_2^r}\frac{i}{2} dz^r\wedge d\bar{z}^r~~;~~ I^{r}=(2\pi R)^2\frac{F_{2r-1\,2r}}{2\pi}\label{bgft2}
\end{eqnarray}
and the same property is true for generalized complex structure as shown in eq. (\ref{gcs}).

The calculus of the Yukawa  couplings for chiral matter involves three different background gauge fields associated to three different stacks of branes. All these  fields
in the complex frame are already $(1,1)$ forms  and their non-vanishing components are  already  diagonal matrices.  The $z$-complex frame  is then universal, being the same for each gauge field.  These latter are  simultaneously diagonal and one can write:
\begin{eqnarray}
I_{ab}+I_{bc}+I_{ca}=0\Rightarrow \lambda^r_{ab}+\lambda^r_{bc}+\lambda^r_{ca}=0~~\forall r=1,2,3 \nonumber
\end{eqnarray}
with $\lambda_r= 2 \pi U_2^r I^{r}/{\cal T}_2^r$, as can be derived from eq. (\ref{bgft2}), and $I\equiv{\rm diag}(I^1,\,I^2,\,I^3)$.

It is now straightforward to evaluate:
\begin{eqnarray}
&&I_{ca}\Omega_{ca}+I_{ab}\Omega_{ab}+I_{bc}\Omega_{bc}= {\rm diag}\left[ \frac{U^r-\bar{U}^r}{2}\left(\eta^r_{ca}I^r_{ca}+\eta^r_{ab}I^r_{ab}+\eta^r_{bc}I^r_{bc}\right)\right]\nonumber\\
&&\Omega_{ab}I_{ab}^{-t}+\Omega_{bc}I_{bc}^{-t}={\rm diag}\left[  \delta_{\eta^r_{ab},\eta^r_{bc}}
(\delta_{\eta^r_{ab},1} U^r+\delta_{\eta^r_{ab},-1}\bar{U}^r) \left( \frac{1}{I^r_{ab}}+\frac{1}{I^r_{bc}}\right)\right.\nonumber\\
&&\left.+\delta_{\eta^r_{ab},-\eta^r_{bc}}\delta_{\eta^r_{ab},1}\left( \frac{U ^r}{I^r_{ab}}+ \frac{\bar{U}^r}{I^r_{bc}}\right)+ \delta_{\eta^r_{ab},-\eta^r_{bc}} \delta_{\eta^r_{ab},-1} \left(\frac{\bar{U}^r}{I^r_{ab}}+ \frac{{U}^r}{I^r_{bc}}\right)  \right]   \nonumber
\end{eqnarray}
with $\eta={\rm sign}\lambda$.
From these expressions one can  compute
\begin{eqnarray}
&&\alpha^{-1}\Pi \alpha^{-t} ={\rm diag} \left[ \left(\frac{1}{I^r_{ab}}+\frac{1}{I^r_{bc}}\right)\left( \left( \delta_{\eta^r_{ab},\eta^r_{bc}}+\delta_{\eta^r_{ab},-\eta^r_{bc}} \delta_{\eta^r_{ab},\eta_r^{ca}}\right) \left(\delta_{\eta^r_{ab},1} U^r+ \delta_{\eta^r_{ab},-1} \bar{U}^r\right)\right.\right.\nonumber\\
&&\left.\left.+ \delta_{\eta^r_{ab},-\eta^r_{bc}} \delta_{\eta_r^{ab},-\eta_r^{ca}}
\left(\delta_{\eta^r_{ab},1} \bar{U}^r+ \delta_{\eta^r_{ab},-1} {U}^r\right)\right)\right] \,\, .  \nonumber
\end{eqnarray}
By using these identities in the Yukawa couplings, an agreement with the corresponding expression, given for example in ref. \cite{0810.5509}, is obtained if $\alpha~=~I_{ab}  I_{bc}$.

\section{Conclusions}
The field theoretical approach pursued in this paper reveals itself to be a very efficient tool in the determination of the low-energy effective actions of branes with oblique magnetization. In particular, it has allowed us to derive the Yukawa couplings for chiral matter in the case of a torus with both an arbitrary magnetization and an arbitrary complex structure. The analogous calculation in a pure stringy approach is still missing, since it has been performed only for a general torus with commuting monodromy  matrices  \cite{0709.1805}.  Of course, this would be an interesting direction to follow, along the lines explored in refs. \cite{1101.5898, 1110.5359} since it could shed light on the string quantization on a general magnetic torus.

Extensions of the results here obtained to the case of magnetization  along the non-Cartan generators of the gauge group could be interesting for exploring connections with F-theory phenomenology.

\vskip 3cm
\begin{center}
{\bf Acknowledgments}
\end{center}

The authors are indebted to Pablo Cam\'ara and Paolo Di Vecchia for stimulating discussions and a critical reading of the manuscript. They thank as well Gaetano Fiore for discussion on related topics.  R. M. thanks Rodolfo Russo for exchanging helpful messages and F. P. thanks Massimo Porrati for useful suggestions.

\appendix
\section{Notations on space-time indices}
The ten-dimensional Minkowski metric is chosen  with the mostly plus signature
\begin{eqnarray}
\eta_{\hat{M}\hat{N}}=(-,\,+,\dots,+)~~;~~\hat{M},\hat{N}=0,\dots, 9\,\, .  \nonumber
\end{eqnarray}
The ten-dimensional indices are separated in the four-dimensional ones $\mu =0,...,3$ relative to the  Minkowski space, and $N=1, \dots, 6$ relative to the compact ones. The compact indices
are called ``flat"  if the coordinates refer to an Euclidean  metric otherwise they are called ``curved".

The capital letters  $M,N=1, \dots,  6$ are used for the real curved coordinates spanning the transverse compact space. The real flat transverse indices are instead denoted by the letters  $I,J=1, \dots, 6$. The indices $m,n=1,\dots, d$ label curved, real or complex,  coordinates while $r,s=1,\dots, d$ denote the corresponding flat ones.

\section{The torus $T^{2d}$}
\label{torus}
The $2d$-dimensional torus $T^{2d}$ is identified with the euclidean space $\mathbb{R}^{2d}$ modulo a $2d$-dimensional lattice $\Lambda=(m^{1} \vec{E}_{1} + m^{2} \vec{E}_{2} +  \dots  + m^{2d} \vec{E}_{2d})$ generated by the set of vectors~ $ (\vec{E}_{1}, \dots, \vec{E}_{2d})$, with $m^{M} \in \mathbb{Z}$.
One can introduce, for the sake of simplicity, the set of versors $ ( \hat{e}_1,\dots, \hat{e}_{2d})$ with $\hat{e}_{M} \equiv \frac{\vec{E}_{M}}{||\vec{E}_{M}||}$. By definition, then, the torus is obtained by imposing on a generic point $P \in \mathbb{R}^{2d}$ of coordinates $\vec{X}$,  the following identification \cite{0810.5509}:
\begin{eqnarray}
X^{I} \equiv X^{I} +2\pi R\, m^M\, e^{I}_{M}  \,\, . \label{ident}
\end{eqnarray}
Let the $2d$ coordinates $X^{M}$ in $\vec{X}$ be grouped in $d$ couples of coordinates $(x^{n}, y^{n})$, according to the definition $(x^{n}, y^{n}) \equiv (X^{2n-1}, X^{2n})$ with $n=1, \dots, d$. Then, one can rewrite eq. (\ref{ident}) in the ``lattice frame'', where the axes are parallel to the versors, as follows:
\begin{eqnarray}
 x^n\equiv x^n +2\pi R\, m^n_1~~;~~y^n\equiv y^n +2\pi R\, m^{n}_2 \,\,   \nonumber
\end{eqnarray}
after having put $m^M =  m^n_1$ $[m^n_{2}]$  for $M=2n-1$ [$M=2n$] with $n=1, \dots, d$. The metric in this frame is given by:
\begin{eqnarray}
ds^2=dX^{M}G_{MN} dX^{N}
\label{rmetric}
\end{eqnarray}
and, denoting by  $\hat{e}^I_{~N}$  the components of the lattice generating  vectors in an orthonormal frame, one gets: $G_{MN}= \hat{e}^I_{~M}\delta_{IJ}\hat{e}^J_{~N}$ ($M,N,I,J=1,\dots, 2d$)\cite{0306006, 0709.1805}. The lattice vectors, by definition, are the vielbein of the metric and allow one to introduce  flat coordinates $(x',\,y')=\hat{e}\cdot X$ having an euclidean   metric.
One can also introduce complex coordinates. The geometry of the complex torus is now described by:
\begin{eqnarray}
d{\cal W}^M={\cal U}^M_{~N} dX^N ~~;~~dX^M=({\cal U}^{-1})^M_{~N} d{\cal W}^N \,\,  \label{complexcoord}
\end{eqnarray}
where  ${\cal W}^M\equiv (w^m,\,\bar{w}^m)$ and
\begin{eqnarray}
{\cal U}=\frac{1}{2\pi R}\left(\begin{array}{cc}
\mathbb{I}&U\\
\mathbb{I}&\bar{U}\end{array}\right)~;~{\cal U}^{-1}=(2\pi R)\left(\begin{array}{cc}
-\frac{\bar{U}}{2i}& \frac{U}{2i} \\
 \frac{\mathbb{I}}{2i}&-\frac{\mathbb{I}}{2i}\end{array}\right)\left(\begin{array}{cc}
                                                                             ({\rm Im}U)^{-1}&0\\
                                                                             0&({\rm Im}U)^{-1}\end{array}\right). \nonumber
\end{eqnarray}
$U$ is the complex structure and the K\"ahler metric is defined by:
\begin{eqnarray}
ds^2=(2\pi R)^2 d\vec{w}^t  \, h  \, d\vec{\bar{w}} \,\, .   \nonumber
\end{eqnarray}
The complex torus is obtained by the identification:
\begin{eqnarray}
w^n=w^n+m^n_1+ U^n_{~p}m^p_2\equiv w^n+ \Pi^n_{~M}m^M   \nonumber
\end{eqnarray}
where the $d \times 2d$ period matrix  $\Pi=(\mathbb{I},\, U)$\cite{complex} has been introduced.
The hermitian metric $h$ can be written in terms of holomorphic and anti-holomorphic vielbeins \cite{Nakahara,0804.0213}
\begin{eqnarray}
h_{m\bar{n}}= e_{~m}^{r}\,\delta_{r \bar{s}}\,\bar{e}_{~\bar{n}}^{\bar{s}}~~;~~r,s=1,\dots, d  \nonumber
\end{eqnarray}
with $(e_m^{~r})^*=\bar{e}_{\bar{m}}^{~\bar{r}}$. Let $e_{~r}^{m}$ and $\bar{e}_{~\bar{s}}^{\bar{n}}$ the inverse of the vielbeins,  i.e. $ \delta_{r\bar{s}}=e_{~r}^{m}h_{m\bar{n}}\bar{e}_{~\bar{s}}^{\bar{n}}$. The vielbeins allow one  to introduce the complex variables:
\begin{eqnarray}
{\cal W}^I=e^I_{~M}{\cal W}^M~~;~~e^I_{~M}=\left(\begin{array}{cc}
                                        e^r_{~m}&0\\
                   0&\bar{e}^{\bar s}_{~\bar{n}}\end{array}\right)   \label{flcomp}
\end{eqnarray}
with $W \equiv(w^r,\,\bar{w}^r)$ having a flat metric
\begin{eqnarray}
ds^2= (2 \pi R)^{2} dw^r\delta_{r \bar{s}}d\bar{w}^{\bar{s}}\equiv \frac{(2 \pi R)^{2}}{2} d\vec{\cal W}^T\,{\cal G}\,d\vec{\cal W} ~~;~~{\cal G}\equiv \left(\begin{array}{cc}
                        0&\mathbb{I}\\
                        \mathbb{I}&0\end{array}\right).  \nonumber
\end{eqnarray}
An orthogonal system of coordinates can be easily introduced by defining new coordinates $(\tilde{x}^r,\,\tilde{y}^r)$
related to the complex flat ones by the relation:
\begin{eqnarray}
X^I= (S^{-1})^I_{~J}{\cal W}^J ~~;~~S=\left(\begin{array}{cc}
                                     \mathbb{I}&i\mathbb{I}\\
                                     \mathbb{I}&-i\mathbb{I}\end{array}\right) \,\, .   \nonumber
\end{eqnarray}
It is trivial to check that $ds^2=(2\pi R)^2 (d\tilde{x}^2+d\tilde{y}^2)$. The relation between  $(\tilde{x},\,\tilde{y})$  and the early quantities $(x,\,y)$ is $X^M= ({\cal U}^{-1})^M_{~N}e^N_{~I}S^I_{~J}X^J\equiv \tilde{e}^M_{~J}X^J$. The quantities denoted by $\tilde{e}$ are the inverse of the vielbein and are mapped by an orthogonal transformation to the vielbein $\hat{e}^M_{~I}$ introduced at the beginning of this appendix.

The ten-dimensional representation of the Clifford algebra
\begin{eqnarray}
\left\{ \Gamma^{\hat{M}},\,\Gamma^{\hat{N}}\right\}=2g^{\hat{M} \hat{N}}  \nonumber
\end{eqnarray}
with $g^{\hat{M}\hat{N}}=(\eta^{\mu\nu},\,G^{MN})$,
is realized  by the following Dirac matrices:\footnote{In the following we restrict our analyses to the case $d=3$.}
\begin{eqnarray}
\Gamma^\mu= \gamma^\mu_{(4)}\otimes \mathbb{I}_{(6)}~~;~~ \Gamma^M=\gamma^5_{(4)}\otimes\gamma^M_{(6)}\label{163}
\end{eqnarray}
with
\begin{eqnarray}
\left\{\gamma^\mu_{(4)},\,\gamma^\nu_{(4)}\right\}=2\eta^{\mu\nu}~~;~~\left\{\gamma^{{M}}_{(6)},\,\gamma^{{N}}_{(6)}\right\}=2G^{{M}{N}} \,\, .  \nonumber
\end{eqnarray}
A representation of the Dirac matrices that satisfies the Clifford algebra with $G^{-1}~= ~2{\cal G}^{-1}$ is
\begin{eqnarray}
\gamma^{w^r}_{(6)}= \mathbb{I}^{\otimes (r-1)}\otimes \gamma^{w^r}
\otimes (\sigma^3)^{\otimes (3-r)}~~;~~
\gamma^{\bar{w}^r}_{(6)}= \mathbb{I}^{\otimes (r-1)}\otimes
\gamma^{\bar{w}^{r}}\otimes (\sigma^3)^{\otimes (3-r)}   \nonumber
\end{eqnarray}
with
\begin{eqnarray}
\gamma^{w^{r}}= \left(\begin{array}{cc}
                                 0&2\\
                                 0&0
                                 \end{array}\right)~~;~~\gamma^{\bar{w}^{r}}=\left(\begin{array}{cc}
                                 0&0\\
                                 2&0
                                 \end{array}\right) \,\, .
\label{curvega}
\end{eqnarray}
It is straightforward to verify that
\begin{eqnarray}
\left\{\gamma^{w^r}_{(6)}\,,\gamma^{{w}^s}_{(6)}\right\}=
\left\{\gamma^{\bar{w}^r}_{(6)}\,,\gamma^{\bar{w}^s}_{(6)}\right\}=0~~;~~  \left\{\gamma^{w^r}_{(6)}\,,\gamma^{\bar{w}^s}_{(6)}\right\}=4\delta^{rs} \,\, .   \nonumber
\end{eqnarray}
The six-dimensional chirality is defined, in the real euclidean system of coordinates, as follows:
\begin{eqnarray}
\gamma_{(6)}^7= -i\prod_{I=1}^6 \gamma_{(6)}^I.  \nonumber
\end{eqnarray}
The $\gamma_{(6)}^I$s satisfy the Clifford algebra $\left\{\gamma_{(6)}^I,\,\gamma_{(6)}^J\right\}=2\delta^{IJ}$ and they are chosen  to be  hermitian matrices.

In the flat complex space the chirality  becomes
\begin{eqnarray}
\gamma_{(6)}^7=-i\prod_{I=1}^6 (S^{-1})^I_{J}\gamma_{(6)}^{J}= \prod_{r=1}^3 \left( \frac{\gamma_{(6)}^{w^r}+\,\gamma_{(6)}^{\bar{w}^r}}{2} \right)
\left( \frac{\gamma_{(6)}^{w^r}-\,\gamma_{(6)}^{\bar{w}^r}}{2} \right) .   \nonumber
\end{eqnarray}
Now, it is simple to convince oneself that a set of eigenfunctions with positive chirality are given by
\begin{eqnarray}
\gamma_{(6)}^{w^1}\gamma_{(6)}^{w^2}\gamma_{(6)}^{w^3}\lambda~~;~~\gamma_{(6)}^{w^1}\gamma_{(6)}^{\bar {w}^2}\gamma_{(6)}^{\bar{w}^3}\lambda\nonumber \\
\gamma_{(6)}^{\bar{w}^1}\gamma_{(6)}^{w^2}\gamma_{(6)}^{\bar{w}^3}\lambda~~;~~
\gamma_{(6)}^{\bar{w}^1}\gamma_{(6)}^{\bar{w}^2}\gamma_{(6)}^{w^3}\lambda   \nonumber
\end{eqnarray}
being $\lambda$ a general eight-dimensional spinor. Notice that in these states it always appears the  product of an odd number of matrices having holomorphic indices. A set of eigenfunctions with negative chirality is obtained by applying on $\lambda$  the product of three Dirac matrices with different space-time indices and with an even number of holomorphic indices:
\begin{eqnarray}
\gamma_{(6)}^{\bar{w}^1}\gamma_{(6)}^{\bar{w}^2}\gamma_{(6)}^{\bar{w}^3}\lambda~~;~~
\gamma_{(6)}^{w^1}\gamma_{(6)}^{w^2}\gamma_{(6)}^{\bar{w}^3}\lambda\nonumber \\
\gamma_{(6)}^{w^1}\gamma_{(6)}^{\bar{w}^2}\gamma_{(6)}^{w^3}\lambda~~;~~
\gamma_{(6)}^{\bar{w}^1}\gamma_{(6)}^{w^2}\gamma_{(6)}^{w^3}\lambda\label{negchir} \,\, .
\end{eqnarray}
In this paper  another set  of complex coordinates, denoted by $({\cal Z}^r,\bar{\cal Z}^r)$, has been  introduced having flat metric and with the property that the background magnetic
field is a block-diagonal  anti-symmetric matrix  (see eq.
(\ref{89})). The Dirac matrices are:
\begin{eqnarray}
\gamma^{{\cal Z}^r}_{(6)}= C^r_{~s} \gamma^{w^s}_{(6)}~~;~~\gamma^{\bar{\cal Z}^r}_{(6)}= \bar{C}^{\bar r}_{~\bar{s}} \gamma^{\bar{w}^s}_{(6)} \,\, ,  \nonumber
\end{eqnarray}
where the unitary matrices $C$ and $\bar{C}$ have been defined in  eq. (\ref{eqei}). Again, in this frame, a set of eigenfunctions with negative chirality is obtained  by applying on a generic spinor the product of an even number of Dirac matrices with holomorphic indices. We can  prove this statement by observing that such a spinor either contains only  anti-holomorphic Gamma matrices or depends on just one anti-holomorphic index,  like the states $\gamma^{{\cal Z}^v}_{(6)}\gamma^{{\cal Z}^w}_{(6)}\gamma^{\bar{\cal Z}^z}_{(6)}\lambda$ ($v\neq w\neq z$). In the first case, one trivially has negative chirality because of eq. (\ref{negchir}), while in the second one, applying $\gamma^7_{(6)}$ on the spinor  yields:
\begin{eqnarray}
&& \gamma^7_{(6)} \,(C^v_{~r}\gamma^{w^r}_{(6)})(C^w_{~s}\gamma^{w^s}_{(6)})(\bar{C}^{\bar z}_{~\bar{t}}
\gamma^{\bar{w}^t}_{(6)})\lambda=- \sum_{r\neq s \neq t=1}^3(C^v_{~r}\gamma^{w^r}_{(6)})(C^w_{~s}\gamma^{w^s}_{(6)})(\bar{C}^{\bar{z}}_{~\bar{t}}
\gamma^{\bar{w}^t}_{(6)})\lambda\nonumber\\
&&\!\!\!+\frac{1}{2^6} \sum_{t\neq r\neq s=1}^3 \!\!\!\left( C^v_{~r} C^w_{~s} \bar{C}^{\bar{z}}_{~\bar{s}}-C^v_{~s}  C^w_{~r} \bar{C}^{\bar{z}}_{~\bar{s}}\right) \gamma^{w^r}_{(6)}\gamma^{\bar{w}^r}_{(6)} \gamma^{w^s}_{(6)}\gamma^{\bar{w}^s}_{(6)} [ \gamma^{w^t}_{(6)}\gamma^{\bar{w}^t}_{(6)}-
\gamma^{\bar{w}^t}_{(6)}\gamma^{w^t}_{(6)} ]
\gamma^{w^r}_{(6)}\gamma^{w^s}_{(6)}\gamma^{\bar{w}^s}_{(6)}\lambda\nonumber\\
&&\!\!\!= -(C^v_{~r}\gamma^{w^r}_{(6)})(C^w_{s}\gamma^{w^s}_{(6)})(\bar{C}^{\bar{z}}_{\bar{t}}
\gamma^{\bar{w}^t}_{(6)})\lambda+\frac{1}{2} \sum_{r\neq s=1}^3 \left( C^v_{~r} C^w_{~s} \bar{C}^z_{~\bar{s}}-C^v_{~s}  C^w_{~r} \bar{C}^{\bar z}_{~\bar{s}}\right)\prod_{u=1}^3\gamma^{w^u}_{(6)}\gamma^{\bar{w}^u}_{(6)} \,\, . \nonumber
\end{eqnarray}
Once that  the identity $\sum_{s=1}^3 C^v_{~s}\bar{C}^{\bar z}_{~\bar{s}}=\delta^{v\bar{z}}$ is used, the last term in the previous equation vanishes. This latter identity follows from the unitarity of the $C^{-1}$ matrices:
\begin{eqnarray*}
(C^{-1})^r_{~p}\delta_{r\bar{s}} (\bar{C}^{-1})^{\bar s}_{~\bar{q}}=\delta_{p\bar{q}}
\end{eqnarray*}
which implies
\begin{eqnarray*}
\delta^{r\bar{s}}C^p_{~r}\delta_{p\bar{q}}= (\bar{C}^{-1})^{\bar{s}}_{~\bar{q}} \Rightarrow \bar{C}^{\bar{t}}_{~\bar{s}} \delta^{r\bar{s}}C^p_{~r}=\delta^{\bar{t}p} .
\end{eqnarray*}

\section{The wave-function} \label{wave-function}
The main ingredient of the bosonic and fermionic wave-function is the Riemann Theta function \cite{MumfordI}:
\begin{eqnarray}
 \mbox{\Large $\Theta$} \mbox{ \scriptsize $ \left[ \begin{array}{c}
\vec{j}\\\vec{i}\end{array}\right]$} (\vec{\nu}|\omega)= \sum_{\vec{n}\in \mathbb{Z}^d} e^{i\pi (\vec{n}+\vec{j})^t \, \omega\,(\vec{n}+\vec{j})+2\pi i (\vec{n}+\vec{j})^t\,(\vec{\nu}+\vec{i})}\label{rtf}
\end{eqnarray}
with $\omega$ being a symmetric $d\times d$ matrix with a positive definite imaginary part. It
satisfies the quasi-periodicity conditions
\begin{eqnarray}
 &&\mbox{\Large $\Theta$} \mbox{ \scriptsize $ \left[ \begin{array}{c}
\vec{j}\\\vec{i}\end{array}\right]$} (\vec{\nu}+\vec{m}|\,\omega)= e^{2\pi i \vec{j}^t \vec{m}}\mbox{\Large $\Theta$} \mbox{ \scriptsize $ \left[ \begin{array}{c}
\vec{j}\\\vec{i}\end{array}\right]$} (\vec{\nu}|\,\omega)\nonumber\\
&& \mbox{\Large $\Theta$} \mbox{ \scriptsize $ \left[ \begin{array}{c}
\vec{j}\\\vec{i}\end{array}\right]$} (\vec{\nu}+\omega \vec{m}|\omega)= e^{-\pi i \vec{m}^t\,\omega \,\vec{m} -2\pi i \vec{m}^t(\vec{\nu}+ \vec{i})}\mbox{\Large $\Theta$} \mbox{ \scriptsize $ \left[ \begin{array}{c}
\vec{j}\\\vec{i}\end{array}\right]$} (\vec{\nu}|\omega) \,\, . \label{qpc}
\end{eqnarray}
It is useful to give the proof of the proposition (6.4) of ref. \cite{MumfordI}, which concerns the product of two Riemann Theta functions
\begin{eqnarray}
&&\! \!\! \! \! \! \! \mbox{\Large $\Theta$} \mbox{ \scriptsize $ \left[ \begin{array}{c}
\frac{\vec{j}_1}{n_1}\\ 0\end{array}\right]$} (\vec{z}_1|n_1\omega)
\mbox{\Large $\Theta$} \mbox{ \scriptsize $ \left[ \begin{array}{c}
\frac{\vec{j}_2}{n_2}\\ 0\end{array}\right]$} (\vec{z}_2|n_2\omega)
=\! \! \! \! \! \! \sum_{\vec{m}\in \mathbb{Z}^d/(n_1+n_2)\mathbb{Z}^d} \mbox{\Large $\Theta$} \mbox{ \scriptsize $ \left[ \begin{array}{c}
\frac{n_1\vec{m}+\vec{j}_1+\vec{j}_2}{n_1+n_2}\\ 0\end{array}\right]$} (\vec{z}_1+\vec{z}_2|(n_1+n_2)\omega)\nonumber\\
 & &\,\,\,\,\,\,  \times \,\,\,\,\,\, \mbox{\Large $\Theta$} \mbox{ \scriptsize $ \left[ \begin{array}{c}
\frac{n_1n_2\vec{m}+n_2\vec{j}_1-n_1\vec{j}_2}{n_1n_2(n_1+n_2)}\\ 0\end{array}\right]$} (n_2\vec{z}_1-n_1\vec{z}_2|n_1n_2(n_1+n_2)\omega)\label{idtheta}
\end{eqnarray}
with $n_1$ and $n_2$ arbitrary integers. Following ref. \cite{MumfordI},
one can introduce the quantities:
\begin{eqnarray}
Q'=\left(\begin{array}{cc}
n_1\omega&0\\
0&n_2\omega\end{array}\right)~~;~~T^{-1}= \left(\begin{array}{cc}
                                                  \mathbb{I}&n_2\mathbb{I}\\
                                                  \mathbb{I}&-n_1\mathbb{I}\end{array}\right)~~;\nonumber\\
Q=T^{-t}Q'T^{-1}= \left(\begin{array}{cc}
(n_1+n_2)\omega&0\\
0&n_1n_2(n_1+n_2)\omega\end{array}\right) \,\, .  \nonumber
\end{eqnarray}
The right side of eq. (\ref{idtheta}) can then be written as follows:
\begin{eqnarray}
\mbox{\Large $\Theta$} \mbox{ \scriptsize $ \left[ \begin{array}{c}
\frac{\vec{j}_1}{n_1}\\ 0\end{array}\right]$} (\vec{z}_1|n_1\omega)
\mbox{\Large $\Theta$} \mbox{ \scriptsize $ \left[ \begin{array}{c}\frac{\vec{j}_2}{n_2}\\ 0\end{array}\right]$} (\vec{z}_2|n_2\omega)&=&\sum_{\vec{l}'\in \mathbb{Z}^{2d}} e^{i\pi (\vec{l}'+\vec{J}')^t \, Q'\, (\vec{l}'+\vec{J}')+ 2\pi i (\vec{l}'+\vec{J}')^t\, \vec{Z}' }
\nonumber\\
&=& \sum_{\vec{l}\in \mathbb{Z}^{2d}} e^{i\pi (\vec{l}+\vec{J})^t \, Q\, (\vec{l}+\vec{J})+ 2\pi i (\vec{l}+\vec{J})^t\, \vec{Z} }\label{lhs}
\end{eqnarray}
with:
\begin{eqnarray}
\vec{Z}= T^{-t}\vec{Z}'= \left(\begin{array}{c}
\vec{z}_1+\vec{z}_2\\n_2\vec{z}_1-n_1\vec{z}_2\end{array}\right) \,\,\,\,\,\,\,& ; & \,\,\,\,\,\,\,  \vec{l}= T \vec{l}'= \left(\begin{array}{c}
\frac{n_1\vec{l}_1+n_2\vec{l}_2}{n_1+n_2} \\  \frac{\vec{l}_1-\vec{l}_2}{n_1+n_2}\end{array}\right) \,\,\,\,\,\,\,;\nonumber\\
\vec{J}=T\vec{J}'  = & & \left(\begin{array}{c}
\frac{\vec{j}_1+\vec{j}_2}{n_1+n_2} \\  \frac{n_2\vec{j}_1-n_1\vec{j}_2}{n_1n_2(n_1+n_2)}\end{array}\right) \,\,  \nonumber
\end{eqnarray}
and
\begin{eqnarray}
\vec{Z}'=\left(\begin{array}{c}
\vec{z}_1\\\vec{z}_2\end{array}\right)~~;~~\vec{l}'=\left(\begin{array}{c}
\vec{l}_1\\ \vec{l}_2\end{array}\right)~~;~~\vec{J}'= \left(\begin{array}{c}
\frac{\vec{j}_1}{n_1}\\ \frac{\vec{j}_2}{n_2}\end{array}\right)  \nonumber
\end{eqnarray}
where the following identities have been used:
\begin{eqnarray}
\frac{n_1\vec{l}_1+n_2\vec{l}_2}{n_1+n_2}= \frac{\vec{m}_1}{n_1+n_2}+\vec{l}_3~~;~~\frac{\vec{l}_1- \vec{l}_2}{n_1+n_2}= \frac{\vec{m}_2}{n_1+n_2}+\vec{l}_4    \label{idsum}
\end{eqnarray}
with $\vec{l}_i\in \mathbb{Z}^d$ ($i=1,\dots, 4$). The quantities $\vec{m}_i$  cannot be any integer  because values of these variables differing by  $\vec{k}(n_1+n_2)$ (with $\vec{k}\in \mathbb{Z}^d$) are equivalent since they determine integer shifts  of the variables $\vec{l}_{3,4}$ which are arbitrary integers.

The quotient space $\mathbb{Z}^d_{n_1+n_2}\equiv\mathbb{Z}^d/[(n_1+n_2)\mathbb{Z}^d]$ is the set of all the inequivalent values of these variables. Furthermore, the $\vec{m}_i$s  cannot be independent because of the following identities:
\begin{eqnarray}
&&\vec{l}_1= \frac{\vec{m}_1+n_2\vec{m}_2}{n_1+n_2}+\vec{l}_3+\vec{l}_4\Rightarrow \vec{m}_1+n_2\vec{m}_2=0~~{\rm mod}(n_1+n_2)\nonumber\\
&&\vec{l}_2= \frac{\vec{m}_1-n_1\vec{m}_2}{n_1+n_2}+\vec{l}_3-\vec{l}_4\Rightarrow \vec{m}_1-n_1\vec{m}_2=0~~{\rm mod}(n_1+n_2) \,\, ,  \nonumber
\end{eqnarray}
which determine $\vec{m}_1=n_1\vec{m}_2$ mod$(n_1+n_2)$. By collecting all these results one can easily get the identity (\ref{idtheta}).

These properties of the Riemann  Theta function are a relevant ingredient  in order to prove that the wave-function introduced in eq. (\ref{ansatz1}) satisfies the boundary conditions given in eq. (\ref{bctorus}). At this aim, one has first to observe that:
\begin{eqnarray}
&&\!\!\!\!\!\phi_0( \vec{\mbox{z}}+ C^{(\lambda)}\eta_{(s)}, \vec{\bar{ \mbox{z}}}+  \bar{C}^{(\lambda)}\eta_{(s)}) = \phi_0(\vec{ \mbox{z}}, \vec{\bar{ \mbox{z}}})
\left[\frac{\theta( \vec{ \mbox{z}}+ C^{(\lambda)}\eta_{(s)})}{\theta(\vec{ \mbox{z}})}\right] e^{\frac{i}{2}
{\rm Im}(\vec{\mbox{z}^t}{C^{(\lambda)}}^{-t}) \bar{C}^{(\lambda)\,t}\,|{\cal I}_{\lambda}|\,C^{(\lambda)}\eta_{(s)}} \, \nonumber\\
&&\,\,\,\,\, \phi_0(\vec{ \mbox{z}}+ C^{(\lambda)}\, \Omega\,\eta_{(s)}, \vec{\bar{ \mbox{z}}}+ \bar{C}^{(\lambda)} \, \bar{\Omega}\,\eta_{(s)}) =\phi_0(\vec{ \mbox{z}}, \vec{\bar{ \mbox{z}}})
\left[\frac{\theta( \vec{ \mbox{z}}+ C^{(\lambda)} \Omega\eta_{(s)})}{\theta(\vec{ \mbox{z}})}\right]\nonumber\\
&&\times \,e^{\frac{i}{2} {\rm Im}(\vec{ \mbox{z}}^t{C^{(\lambda)}}^{-t}) \bar{C}^{(\lambda)\,t} |{\cal I}_\lambda| C^{(\lambda)}\, \Omega\,\eta_{(s)}+  \frac{i}{2} \eta_{(s)}^t\,{\rm Im}\Omega^t \,\bar{C}^{(\lambda)\,t} |{\cal I}_\lambda| \vec{ \mbox{z}}
+ \frac{i}{2} \eta_{(s)}^t\,{\rm Im}\Omega^t \,\bar{C}^{(\lambda)\,t} |{\cal I}_\lambda|C^{(\lambda)}\, \Omega\,\eta_{(s)}}\label{imid}
\end{eqnarray}
where $\Omega$ is defined after eq.(\ref{id2}) and the identity $  {\bar C}^{(\lambda)\,t}  |{\cal I}_{\lambda}| C^{(\lambda)}={C}^{(\lambda)\,t} |{\cal I}_{\lambda}| \bar{C}^{(\lambda)}$ has been used.

By plugging the first of the two last identities in eq.  (\ref{bctorus}), one gets the first identity written in eq. (\ref{IIbc})
while, by plugging in the second line of  eq. (\ref{bctorus}) the last identity in eq. (\ref{imid})
and using again the symmetry of the matrix ${\bar C}^{(\lambda)\,t}  |{\cal I}_{\lambda}| C^{(\lambda)}$, one gets:
\begin{eqnarray}
\theta(  \vec{\mbox{z}}+ C^{(\lambda)} \Omega \eta_{(s)})= e^{- \frac{i}{2}\eta_{(s)} {\rm Im}\Omega^t {C^{(\lambda)}}^t| {\cal I}_\lambda|\bar{C}^{(\lambda)}[2{C^{(\lambda)}}^{-1}\vec{ \mbox{z}} +\Omega\,\eta_{(s)}]}
\theta( \vec{ \mbox{z}}) \,\, .   \nonumber
\end{eqnarray}
It is worth to stress here that, in writing these boundary conditions,  the matrix
$C^{(\lambda)}$ has been implicitly assumed to be invertible.

The complex coordinates are related to the real ones by the identities:
\begin{eqnarray}
\vec{\mbox{z}}=\frac{C^{(\lambda)} \vec{x}+ C^{(\lambda)} \Omega \vec{y}}{2\pi R}~~;~~\vec{\bar{\mbox{z}}}=\frac{\bar{C}^{(\lambda)} \vec{x}+ \bar{C}^{(\lambda)} \bar{\Omega}  \vec{y}}{2\pi R} \,\, . \nonumber
\end{eqnarray}
The magnetic background  flux in both systems of coordinates can be read from eq.~ (\ref{1}) and eq. (\ref{29}), here rewritten:
\begin{eqnarray*}
F&=&\frac{i}{4} d\vec{\mbox{z}}^t  |{\cal I}_{\lambda}| \wedge d\vec{\bar{\mbox{z}}}-\frac{i}{4} d\vec{\bar{\mbox{z}}}^t  |{\cal I}_{\lambda}| \wedge d\vec{\mbox{z}}\nonumber\\
&=& \frac{1}{2}d\vec{x}^t  F^{(xx)} \wedge d\vec{x}+ \frac{1}{2}d\vec{x}^t F^{(xy)} \wedge d\vec{y}-
\frac{1}{2}d\vec{y}^t  {F^{(xy)t}} \wedge d\vec{x}+ \frac{1}{2}d\vec{y}^t F^{(yy)} \wedge d\vec{y}
\end{eqnarray*}
with
\begin{eqnarray}
F^{(xx)}&=&\frac{i}{2(2\pi R)^2}\left[ {C^{(\lambda)}}^t| {\cal I}_\lambda| \bar{C}^{(\lambda)}- \bar{C}^{(\lambda)\,t}|\mathbb{I}_{\lambda}| C^{(\lambda)}\right]\nonumber\\
F^{(xy)}&=&\frac{i}{2(2\pi R)^2}\left[ {C^{(\lambda)}}^t|\mathbb{I}_\lambda| \bar{ C}^{(\lambda)}\, \bar{\Omega}- \bar{C}^{(\lambda)\,t}| {\cal I}_{\lambda}|{C}^{(\lambda)}\,\Omega\right]\nonumber\\
-{F^{(xy)}}^t&=& \frac{i}{2(2\pi R)^2}\left[\Omega^t\,{ C}^{(\lambda)\,t}| {\cal I}_\lambda| \bar{C}^{(\lambda)}-\bar{\Omega}^t\, \bar { C}^{(\lambda)\,t}| {\cal I}_{\lambda}| {C}^{(\lambda)}\right]\nonumber\\
F^{(yy)}&=& \frac{i}{2(2\pi R)^2}\left[\Omega^t \,{C}^{(\lambda)\,t}| {\cal I}_\lambda|\bar{ C}^{(\lambda)}\,\bar{\Omega}-\bar{\Omega}^t\,\bar{ C}^{(\lambda)\,t}| {\cal I}_\lambda|\,{C}^{(\lambda)}\,\Omega\right]   \,\, .\label{70}
\end{eqnarray}
It is now straightforward to see that requiring the matrix  ${\bar C}^{(\lambda)\,t}  |\mathbb{I}_{\lambda}| C^{(\lambda)}$ to be symmetric implies the first identity written in eq. (\ref{148}), $F^{(xx)}=0$ and allows one to write:
\begin{eqnarray}
F^{(yy)}&=& -\frac{i}{2(2\pi R)^2} \bar{\Omega}^t \bar{C}^{(\lambda)\,t}| {\cal I}_\lambda|C^{(\lambda)}\Omega +\Omega^t F^{(xy)}+ \frac{i}{2(2\pi
R)^2}\Omega^t \bar{C}^{(\lambda)\,t}| {\cal I}_\lambda| {C}^{(\lambda)}{\Omega}\nonumber\\
&=& -\frac{1}{(2\pi R)^2}{\rm Im}\Omega^t \bar{C}^{(\lambda)\,t}| {\cal I}_\lambda|C^{(\lambda)}\Omega +\Omega^t F^{(xy)}=-{F^{(xy)}}^t\Omega+\Omega^tF^{(xy)}  \nonumber
\end{eqnarray}
where  the second line of eq. (\ref{70}) has been used.

The second line of eq. (\ref{fbc}) becomes identical to the second identity in eq. (\ref{qpc}) if the quantity  $I\, \Omega$, with $I= (2\pi R)^2\frac{{F^{(xy)}}^t}{2\pi}$ is identified with the symmetric matrix $\omega$, if $\vec{b}$ vanishes and if $\vec{\nu}= I  {C^{(\lambda)}}^{-1} \mbox{z}$. These identifications lead to the solution written in eq. (\ref{sol}).

In order to verify if the Riemann Theta function satisfies also the first line of eq. (\ref{fbc}), one explicitly computes:
\begin{eqnarray}
\Theta\left[\begin{array}{c}\vec{j}\\0\end{array}\right](I\, (C^{(\lambda)\,-1} \vec{Z}+\eta_m)| I\, \Omega)&=&
\sum_{\vec{n}\in \mathbb{Z}^d}e^{i\pi(\vec{n}+ \vec{j})^t \,[ I\, \Omega]\,(\vec{n}+\vec{j})+2\pi i (\vec{n}+ \vec{j})^t  \, I\, C^{(\lambda)\,-1} \vec{\mbox{z}}   }\nonumber\\
&\times& e^{2\pi i (\vec{n}+ \vec{j})^t  \, I\,\eta_{(s)}}\nonumber
\end{eqnarray}
and imposes the additional constraints:
\begin{eqnarray}
(\vec{n}^t\, I)\in \mathbb{Z}^{d}~~;~~  (\vec{j}^t\, I) \in \mathbb{Z}^{d}~~~  \forall \vec{n}\in \mathbb{Z}^{d}.  \nonumber
\end{eqnarray}
The first constraint is in fact satisfied because $I$ is a matrix with integer entries.

In the last part of this appendix, the consistency of the solution (\ref{83}) is going to be proved.
The wave-function is correctly defined if the Riemann Theta function is a convergent series and the  convergence is guaranteed if the symmetric matrix  $I\, {\rm Im}\Omega$ is positive definite~\cite{MumfordI}, i.e.:
\begin{eqnarray}
\vec{x}^{~t}\, (2\pi R)^2\frac{{F^{(xy)}}^t}{2\pi}\, {\rm Im} \Omega\,\vec{x}> 0 ~~;~~\forall \vec{x} \in \mathbb{R}^d~~{\rm and}~~\vec{x}\neq 0 \,\, .  \nonumber
\end{eqnarray}
At this aim one has first to observe that eq. (\ref{148})  implies:
\begin{eqnarray*}
\vec{x}^{~t} \,(2\pi R)^2\frac{F^{(xy)}}{\pi}\, {\rm Im} \Omega^{-1}\,\vec{x}=
(\vec{x} C^{(\lambda)})^\dagger \, \frac{|{\cal I}_\lambda|}{\pi} \, ( C^{(\lambda)}\vec{x})=\sum_m |\frac{\lambda_r}{\pi}||(\vec{x}C^{(\lambda)})_m|^2 >0~~.
\end{eqnarray*}
Furthermore one has:
\begin{eqnarray}
\vec{x}^t \,[{F^{(xy)}}^t \, {\rm Im}\Omega]\, \vec{x} & = & \vec{x}^T\,{F^{(xy)}}^t  \, {\rm Im}\Omega\, {\rm Im}\Omega^{-1} \, {\rm Im}\Omega \, \vec{x} \nonumber \\
&= & \vec{x}^t\,{\rm Im}\Omega^t {F^{(xy)}}  \,{\rm Im}\Omega^{-1} \,{\rm Im}\Omega\,\vec{x}\nonumber\\
&= &({\rm Im}\Omega\,\vec{x})^t\, [ {F^{(xy)}}  {\rm Im}\,\Omega^{-1}]\,(\Omega\,\vec{x})\equiv \vec{v}^t [ {F^{(xy)}}  {\rm Im}\,\Omega^{-1}]\vec{v}>0 \nonumber
\end{eqnarray}
since the matrix $[{F^{(xy)}} {\rm Im}\Omega^{-1}]$ is positive
definite. This proves the convergence of the Theta function.
In the proof  the following identity has been used:
\begin{eqnarray*}
{F^{(xy)}}^t\Omega=\Omega^tF^{(xy)}\Rightarrow {F^{(xy)}}^t{\rm Im}\Omega={\rm Im}\Omega^tF^{(xy)} \,\, .
\end{eqnarray*}
The normalization of the wave-function is determined by evaluating the integral:
\begin{eqnarray*}
&&\frac{1}{2g}\int d^dx\,d^dy \sqrt{G_6} (\phi^{ab}_j)^*\, \phi^{ab}_{j'}=\frac{{\cal C}^2}{2g} \sqrt{G_6} \sum_{\vec{n},\vec{m}\in\mathbb{Z}^d} \int_0^{2\pi R}d^dx \,
e^{2\,i\, \pi (\vec{n}+\vec{j}-\vec{m}-\vec{j}')\frac{I_{ab}}{2\pi R} \vec{x}}\nonumber\\
&& \!\!\!\! \times \, e^{i\pi (\vec{n}+\vec{j})^t I_{ab} \Omega_{ab} (\vec{n}+\vec{j})-i\pi (\vec{m}+\vec{j}')I_{ab}\bar{\Omega}_{ab}(\vec{m}+\vec{j}')} \int_0^{2\pi R} d^dy e^{i\pi \vec{y}^t(\Omega_{ab}-\bar{\Omega}_{ab}) \vec{y}} e^{2i\pi [(\vec{n} +\vec{j})
-(\vec{m}+\vec{j}')]\frac{I_{ab}}{2\pi R} (\Omega_{ab}-\bar{\Omega}_{ab}) }. \nonumber
\end{eqnarray*}
The integral over the $x$-coordinates is trivial, giving the result:
\begin{eqnarray*}
\int_0^{2\pi R} \, d^dx \,\,
e^{2\,i\, \pi (\vec{n}+\vec{j}-\vec{m}-\vec{j}')\frac{I_{ab}}{2\pi R} \vec{x}}=(2\pi R)^d \delta^{(d)}_{\vec{n}+\vec{j},\vec{m}+\vec{j}'}
\end{eqnarray*}
which imposes $\vec{n}=\vec{m}$ and $\vec{j}=\vec{j}'$. Using these latter identities we get:
\begin{eqnarray*}
\!\!\!\!\!\!\!&&\frac{1}{2g}\int d^dx\,d^dy \sqrt{G_6} (\phi^{ab}_j)^*\, \phi^{ab}_{j'}=\delta_{\vec{j},\vec{j}'}^{(d)}\frac{{\cal C}^2}{2g} (2\pi R)^{2d} \sqrt{G_6} \sum_{\vec{n}\in\mathbb{Z}^d}\int_0^1d^d y e^{-2\pi (\vec{y}+\vec{n} +\vec{j})^tI_{ab}{\rm Im}\Omega_{ab} (\vec{y}+\vec{n} +\vec{j})} \,\, .
\end{eqnarray*}
The last integral in the previous equation is evaluated by observing that:
\begin{eqnarray*}
&&\sum_{\vec{n}\in\mathbb{Z}^d}\int_0^1d^d y e^{-2\pi (\vec{y}+\vec{n} +\vec{j})^tI_{ab}{\rm Im}\Omega_{ab} (\vec{y}+\vec{n} +\vec{j})}=\prod_{i=1}^d \left[\lim_{A^i\rightarrow \infty}\sum_{n^i=-A^i}^{A^i}
\int_{n^i+j^i}^{n^i+j^i+1}dy^i\right]e^{-2\pi \vec{y}^tI_{ab}{\rm Im}\Omega_{ab}\vec{y}}\nonumber\\
&&= \prod_{i=1}^d \left[\lim_{A^i\rightarrow \infty} \int_{-A^i}^{A^i}dy^i\right]
e^{-2\pi \vec{y}^tI_{ab}{\rm Im}\Omega_{ab}\vec{y}}=\int_{-\infty}^{\infty} d^dy e^{-2\pi \vec{y}^tI_{ab}{\rm Im}\Omega_{ab}\vec{y}}\nonumber\\
&&= [{\rm det}(I_{ab}{\rm Im}\Omega_{ab})]^{-1/2}
\end{eqnarray*}
Hence, the wave-function normalization turns out to be:
\begin{eqnarray}
\frac{1}{2g}\int d^dx\,d^dy \sqrt{G_6} (\phi^{ab}_j)^*\, \phi^{ab}_{j'}=    \frac{{\cal N}^2}{2g} V_{T^{2d}}\delta^{(d)}_{\vec{j},\vec{j}'}
[{\rm det}( I_{ab}{\rm Im}\Omega_{ab})]^{-1/2}  \nonumber
\end{eqnarray}
with $V_{T^{2d}}=(2\pi R)^{2d} \sqrt{G_6}$.
Choosing ${\cal C}=\sqrt{2g}V_{T^{2d}}^{-1/2} \left[ {\rm det}( I_{ab}{\rm Im}\Omega_{ab})\right]^{1/4}$  makes  the kinetic term of the scalars  canonically normalized.

\section{Explicit calculus of the Yukawa couplings}
\label{Yukawa calculus}

The Yukawa couplings involving two twisted fermions and one twisted scalar field,
corresponding to the lowest excitations of the dy-charged open strings ending on three magnetized branes, are obtained by computing the following integral:
\begin{eqnarray}
{\cal Y}^{\vec{j}_1,\,\vec{j}_2,\,\vec{j}_3}= \int_{T^6} d^3\tilde{x} d^3 \tilde{y}\sqrt{G_6}(\phi^{ac}_{\vec{j}_3}(\Omega_{ac}))^*\,
\phi^{\,ab}_{\vec{j}_1}(\Omega_{ab})\, \phi^{bc}_{\vec{j}_2}(\Omega_{bc})  \nonumber
\end{eqnarray}
being $(\vec{\tilde{x}},\,\vec{\tilde{y}})=(\vec{x},\,\vec{y})/(2\pi R)$ and  the wave-function, defined in eq. (\ref{83}), is  here rewritten:
\begin{eqnarray}
\phi_{\vec{j}}(\Omega) = {\cal N} e^{i\pi \vec{\tilde{y}}^t  I  \vec{\tilde{x}}+i\pi  \vec{\tilde{y}}^t  I\Omega \vec{\tilde{y}}}\sum_{\vec{n}\in \mathbb{Z}^{3}}
e^{ i\pi  (\vec{n}+\vec{j})^t   I \Omega(\vec{n}+\vec{j})+ 2 \pi i  (\vec{n}+\vec{j})^t\, I (\vec{\tilde{x}}+\Omega\, \vec{\tilde{y}})}\label{83b}
\end{eqnarray}
with $I= (2\pi R)^2\,F^{(xy)t}/2 \pi$.

It is useful to give to the product of two wave-functions the following form \cite{0904.0910}:
\begin{eqnarray}
\phi_{\vec{j}_1}^{ab}(\Omega_{ab})\, \phi_{\vec{j}_2}^{bc}(\Omega_{bc})&=&{\cal N}_{ab}\,{\cal N}_{bc} \,e^{i\pi \vec{\tilde y}^t\,(I_{ab}+I_{bc})^t \, \vec{\tilde x}+i\pi  \vec{\tilde y}^t (I_{ab}\,\Omega_{ab}+I_{bc}\,\Omega_{bc}) \vec{\tilde y}}\nonumber\\
&\times&\sum_{\vec{l}\in \mathbb{Z}^{2d}}e^{i\pi \vec{l}^t  Q \, \vec{l}+2\pi i \vec{l}^t  Q  \vec{\tilde Y}
+ 2\pi i \vec{l}^t {\cal I}  \vec{\tilde X}}   \nonumber
\end{eqnarray}
being
\begin{eqnarray}
Q=\left(\begin{array}{cc}
         I_{ab}\Omega_{ab} &0\\
         0& I_{bc}\Omega_{bc}\end{array}\right)~~&;&~~ {\cal I}=\left(\begin{array}{cc}
                                                               { I}_{ab}&0\\
                                                               0&{ I}_{bc}\end{array}\right)\nonumber\\
\vec{l}=\left(\begin{array}{c}
\vec{n}_1+\vec{j}_1\\
\vec{n}_2+\vec{j}_2\end{array}\right)~~;~~\tilde{X}=\left(\begin{array}{c}
                        \vec{\tilde x}\\
                        \vec{\tilde x}\end{array}\right)~~&;&~~\tilde{Y}=\left(\begin{array}{c}
                        \vec{\tilde y}\\
                        \vec{\tilde y}\end{array}\right) \,\,\, .    \nonumber
\end{eqnarray}
An equivalent representation of the product of two Riemann Theta functions is obtained by introducing the following transformation matrix:
\begin{eqnarray}
T=\left(\begin{array}{cc}
        \mathbb{I}&\mathbb{I}\\
        \alpha I_{ab}^{-1}&-\alpha I_{bc}^{-1}\end{array}\right) ~~;~~
        T^{-1}= \left( \begin{array}{cc}
        (I_{ab}^{-1}+I_{bc}^{-1})^{-1}
        I_{bc}^{-1} &(I_{ab}^{-1}+I_{bc}^{-1})^{-1} \alpha^{-1}\\
        (I_{ab}^{-1}+I_{bc}^{-1})^{-1} I_{ab}^{-1} &-(I_{ab}^{-1}+I_{bc}^{-1})^{-1}\alpha^{-1}\end{array}\right) \nonumber
\end{eqnarray}
acting as follows:
\begin{eqnarray}
&&Q'=T  Q  T^t= \left(\begin{array}{cc} Q'^{11} & Q'^{12} \\ Q'^{21} & Q'^{22} \end{array}  \right) =
\left(\begin{array}{cc}
                           I_{ab}\Omega_{ab}+I_{bc}\Omega_{bc}& (\Omega_{ab}^t-\Omega_{bc}^t)\alpha^t\\
                           \alpha(\Omega_{ab}-\Omega_{bc})& \alpha(\Omega_{ab}I_{ab}^{-t}+\Omega_{bc}I_{bc}^{-t})\alpha^t\end{array}\right)\nonumber\\
&&{\cal I}'= T\, {\cal I}\, T^t=\left(\begin{array}{cc}
                              I_{ab}+I_{bc}& (I_{ab}\,I_{ab}^{-t}-I_{bc}\,I_{bc}^{-t})\alpha^t\\
                             0& \alpha( I_{ab}^{-t}+I_{bc}^{-t})\alpha^t
                              \end{array}\right) \,\, .    \nonumber
\end{eqnarray}
We introduce also the vector:
\begin{eqnarray}
\vec{l}^{\, t}T^{-1}&=& \left( (\vec{n}_1+\vec{j}_1)^t(I_{ab}^{-1}+I_{bc}^{-1})^{-1}I_{bc}^{-1} + (\vec{n}_2+\vec{j}_2)^t(I_{ab}^{-1}+I_{bc}^{-1})^{-1}I_{ab}^{-1}\,;\right.\nonumber\\
&& \left.(\vec{n}_1+\vec{j}_1)^t(I_{ab}^{-1}+I_{bc}^{-1})^{-1}\alpha^{-1}
- (\vec{n}_2+\vec{j}_2)^t(I_{ab}^{-1}+I_{bc}^{-1})^{-1}\alpha^{-1}\right)  \,\, .  \nonumber
\end{eqnarray}
By using the following identity:
\begin{eqnarray}
\left( I_{ab}^{-1}+I_{bc}^{-1}\right)^{-1}= I_{bc}\left(I_{ab}+I_{bc}\right)^{-1}I_{ab}= I_{ab}\left(I_{ab}+I_{bc}\right)^{-1}I_{bc}  \nonumber
\end{eqnarray}
and analogously to what has been done in  eqs.  (\ref{idsum}), one can write:
\begin{eqnarray}
&&\left(\vec{n}_1^t \,I_{ab}+\vec{n}_2^t\,I_{bc}\right)\left(I_{ab}+I_{bc}\right)^{-1}= \vec{m}_1^t \left(I_{ab} +I_{bc} \right)^{-1}+\vec{l}_3^t \nonumber \\
&& \left( \vec{n}_1^t-\vec{n}_2^t \right)I_{ab}\left(I_{ab}+I_{bc}\right)^{-1} I_{bc}\alpha^{-1}= \vec{m}_2^t \left(I_{ab}^{-1}+I_{bc}^{-1}\right)^{-1} \alpha^{-1}+\vec{l}_4^t\label{mdef}
\end{eqnarray}
where $\vec{l}_3,\,\vec{l}_4\in \mathbb{Z}^3$, $\vec{m}_1$ and $\vec{m}_2$ are suitable  integer vectors, while $\alpha$ has to be chosen in  such a way  that the matrix $ \alpha \left(I_{ab}^{-1}+I_{bc}^{-1}\right) $ has integer entries.  In the following, we will choose  $\alpha= {\rm det} \left[ I_{ab}I_{bc}\right] \mathbb{I}$ \cite{0904.0910} which indeed satisfies the above mentioned constraint. The possible values of $\vec{m}_{1,2}$ can be determined by  repeating the analysis  developed after eq. (\ref{aparameter}). By writing  $\vec{m}_1= m_1^i\vec{e}_i$,  the lattice with basis vectors $\vec{e}_i \left(I_{ab}+I_{bc}\right)$ is introduced  and, in it,  the equivalent points are those which change $\vec{l}_3$ by   integer values, because this quantity is summed over all possible elements of $\mathbb{Z}^3$.

$\mathbb Z^{3}_{(I_{ab}+I_{bc})}$ is the set of equivalent classes obtained by identifying the elements of ${\mathbb{Z}^3}$  under the shift $\vec{m}_1+\vec{k}^t\left(I_{ab}+I_{bc}\right)$ ($\forall \vec{k}\in \mathbb{Z}^3$). Inequivalent values of $\vec{m}_1$ lie in the cell determined by the vectors $\vec{e}_i \left(I_{ab}+I_{bc}\right)$,
their number is $|{\rm det} [I_{ab}+I_{bc}]|$.  Analogously,  the number of inequivalent values of $\vec{m}_2 \in \mathbb Z^{3}_{(I_{ab}^{-1}+I_{bc}^{-1})\alpha}$  is
$|{\rm det }[I_{ab}^{-1}+I_{bc}^{-1}]\alpha|$.

It is  straightforward to obtain from the previous equations the identities:
\begin{eqnarray}
&&\vec{n}_1^t= (\vec{m}_1^t+\vec{m}_2^tI_{bc})(I_{ab}+I_{bc})^{-1}+\vec{l}_3^{\, t}+\vec{l}_4^{\, t} \alpha I_{ab}^{-1}
\nonumber\\
&&\vec{n}_{2}^t=(\vec{m}_1^t-\vec{m}_2^tI_{ab})(I_{ab}+I_{bc})^{-1} +\vec{l}_3-\vec{l}_4\alpha I_{bc}^{-1} \,\, . \nonumber
\end{eqnarray}
In order to make them consistent,  both $\alpha I^{-1}_{ab}$ and $\alpha I_{bc}^{-1}$  have to be integers and this  is  satisfied by choosing  $\alpha={\rm det}[I_{ab}\,I_{bc}]\mathbb{I}$.\footnote{The most general solution would be   $\alpha={\rm det}[I_{ab}\,I_{bc}]P$ with $P$ an integer matrix\cite{0904.0910}.} Moreover, one has to impose also:
\begin{eqnarray}
\vec{m}_1^t+\vec{m}_2^tI_{bc}=\vec{k}^{t} (I_{ab}+I_{bc}) ~~;~~\vec{m}_1^t-\vec{m}_2^tI_{ab}=\vec{k}_{1}^{t}(I_{ab}+I_{bc})   \nonumber
\end{eqnarray}
with $\vec{k}$ and $\vec{k}_1$ elements of $\mathbb{Z}^3$. The solution of the last two equations is
\begin{eqnarray}
\vec{m}_1^t=\vec{m}_2^tI_{ab}+\vec{k}_1^{t}(I_{ab}+I_{bc}) \,  .\label{modi}
\end{eqnarray}
Eq. (\ref{modi}), after having  taken into  account the different definitions of the equivalence classes associated to the two integer vectors $\vec{m}_{1,2}$, becomes:
\begin{eqnarray}
\vec{m}_1^t+ \vec{t}_1^t(I_{ab}+I_{bc})=\left[ \vec{m}_2^t+\vec{t}_2^t\alpha (I_{ab}^{-1}+I_{bc}^{-1})\right]I_{ab}+\vec{k}_{1}^{t}(I_{ab}+I_{bc}) .  \nonumber
\end{eqnarray}
This is equivalent to eq. (\ref{modi}) with $\vec{k}_1^{t}$ replaced by
$\vec{k}_1^{t}-\vec{t}_1^{\, t}+\vec{t}_2^{\, t}\alpha I_{bc}^{-1}$.
The identification written in eq. (\ref{modi}) is consistent because the vectors $\vec{m}_1$ and $\vec{m}_2I_{ab}$ are both identified up to the integer matrix  $(I_{ab}+I_{bc})$, being
\begin{eqnarray}
\vec{m}_2^tI_{ab}\equiv\vec{m}_2^tI_{ab}+ (\vec{k}^t\,I_{bc}^{-1} {\rm det}[I_{ab}I_{bc}])(I_{ab}+I_{bc}). \nonumber
\end{eqnarray}
The correspondence between $\vec{m}_1$ and $\vec{m}_2$ is not one-to-one
since, having been chosen $\alpha=~\mbox{det}[I_{ab}I_{bc}]\mathbb{I}$, the number of the inequivalent values of $\vec{m}_2$  is bigger than the one of inequivalent $\vec{m}_1$. Following ref. \cite{0904.0910}, one can  replace:
\begin{eqnarray}
\vec{m}_{2}^{t}=\vec{\tilde m}_{2}^{t}+\vec{p}^{t} {\rm det}[I_{ab}] (I_{ab}+I_{bc}) I_{ab}^{-1}+ \vec{\tilde p}^{t} \, {\rm det}[I_{bc}] (I_{ab}+I_{bc}) I_{bc}^{-1}\label{mtildem}
\end{eqnarray}
and the second line of eq. (\ref{mdef}) becomes:
\begin{eqnarray}
\left( \vec{n}_1^t-\vec{n}_2^t \right)I_{ab}\left(I_{ab}+I_{bc}\right)^{-1} I_{bc}\alpha^{-1} \, \, & =  & \,\,  \vec{\tilde m}_2^t \left(I_{ab}^{-1}+I_{bc}^{-1}\right)^{-1} \alpha^{-1}+\vec{p}^{\,t}\frac{I_{bc}}{{\rm det}I_{bc}} \nonumber
\\ &+ &   \vec{\tilde p}^{\, t} \frac{I_{ab}}{{\rm det}I_{ab}}+  \vec{l}_4^{\, t} . \label{newid}
\end{eqnarray}
The sets of  inequivalent $\vec{p}$ and $\vec{\tilde p}$ are respectively denoted by $\mathbb Z^{3}_{{\rm det}[I_{bc}] I_{bc}^{-1} }$ and $\mathbb Z^{3}_{{\rm det}[I_{ab}] I_{ab}^{-1} }$ and their numbers are
$|{\rm det}({\rm det }[I_{bc} ] I_{bc}^{-1})|$ and $|{\rm det}({\rm det }[I_{ab}] I_{ab})^{-1}|$. Consequently, the number of inequivalent $\vec{\tilde m}_2$'s is $|{\rm det} (I_{ab}+I_{bc})|$ which now matches with the one of the $\vec{m}_{1}$'s. It is worthwhile to observe that the previous counting  is correct only if the Chern classes $I_{ab}$ and $I_{bc}$ are completely independent. When one of these two conditions $I_{ab}=P\,I_{bc}$ or $I_{bc}=P\,I_{ab}$, with $P$  a matrix with integer entries, is verified - for example the second identity -  then:
\begin{eqnarray}
\frac{I_{ab}}{{\rm det}[I_{ab}]}=({\rm det}[P] P^{-1}) \frac{I_{bc}}{{\rm det}[I_{bc}]}\label{newid1} \,\,\, ,
\end{eqnarray}
being ${\rm det}[P] P^{-1}$ a matrix with integer entries. The right side of eq. (\ref{newid}) becomes:
\begin{eqnarray}
\vec{\tilde m}_{2}^{t}(I_{ab}^{-1}+I_{bc}^{-1})^{-1}\alpha^{-1}+( \vec{ p}^{t}+\vec{\tilde p}^t {\rm det}[P]P^{-1}) \, \frac{I_{bc}}{{\rm det}[I_{bc}]} \,\, . \label{mtildem1}
\end{eqnarray}
The latter equation shows that the sum over $\vec{\tilde p}$ introduces  integers already taken into account in the sum over $\vec{p}$. The overcounting is $|{\rm det}[{\rm det}[I_{ab}]I_{ab}^{-1}]|$ and the sum over these two integers has to be normalized by this factor for avoiding to include  equivalent contributions several times. Furthermore, in this case the degeneracy of $\vec{\tilde{m}}$ is bigger exactly by the factor $|{\rm det}[{\rm det}[I_{ab}]I_{ab}^{-1}]|$.
Analogous considerations are valid when the other condition $I_{ab}=P\,I_{bc}$ is verified.

By  starting from eq. (\ref{newid}) and repeating the same manipulation which has led to eq. ({\ref{modi}), one gets  the same equation with $\vec{m}_2$ replaced by  $\vec{\tilde m}_2$. The solution of eq. (\ref{modi}), in the case of $I_{ab}$ and $I_{bc}$ independent,  is now unique and  one can write ($\vec{m}\equiv \vec{\tilde m}_2$):
\begin{eqnarray}
&&\vec{l}^{~' t}=(\vec{l}_{1}^{~'t},\, \vec{l}_2^{~' t})\equiv \vec{l}^{\, t}T^{-1}\equiv \left( (\vec{j}_1^{t} I_{ab} +\vec{j}_2^{t}I_{bc} +\vec{m}^t I_{ab})\left(I_{ab} +I_{bc} \right)^{-1}+\vec{l}_3^{\, t};\right.\nonumber\\
&&\left.(\vec{j}_1^t- \vec{j}_2^t+\vec{ m}^t) I_{ab} \left(I_{ab} +I_{bc} \right)^{-1}I_{bc} \alpha^{-1}+ \vec{p}^{\, t}\frac{I_{bc}}{{\rm det}[I_{bc}]} +  \vec{\tilde p}^{\,t}\frac{I_{ab}}{{\rm det}[I_{ab}]}+   \vec{l}_4^{\, t} \right) . \nonumber
\end{eqnarray}
The case  in which $I_{ab}$ and $I_{bc}$ are not independent, for example $I_{bc}=P\,I_{ab}$, is subtle  because the degeneracy of $\vec{\tilde m}_2$ is bigger than the one of $\vec{m}_1$. Furthermore, in order to satisfy the condition:
\begin{eqnarray}
(\vec{m}_1- \vec{\tilde m}_2I_{ab})(I_{ab}+I_{bc})^{-1}=0 ~~{\rm mod}\, \mathbb{Z}^d  \nonumber
\end{eqnarray}
one has to find the number of inequivalent values of  $\vec{\tilde m}_2$ according to the identification
\begin{eqnarray}
\vec{\tilde m}_2\equiv \vec{\tilde m}_2+ \vec{k}^t(1+P) \,\, .  \nonumber
\end{eqnarray}
This number is $|{\rm det}(1+P)|$
which is smaller than the number of possible $\vec{m}_1$. By remembering that the degeneracy of $\vec{\tilde m}_2$ is $|{\rm det}[({\rm det}[I_{ab}]I_{ab}^{-1})(I_{ab}+I_{bc})]|$, the number of times that inequivalent values of $\vec{\tilde m}_2$ appear is $|{\rm det}[I_{ab}]|^{d}$.

From the above analysis immediately it  follows that:
\begin{eqnarray}
\vec{\tilde m}_2\in \tilde{\mathbb{Z}}_{(I_{ab}^{-1}+I_{bc}^{-1})\alpha}=\mathbb{Z}^3_{(I_{ab}^{-1}+I_{bc}^{-1})\alpha}\setminus (\mathbb Z^{3}_{{\rm det}[I_{bc}] I_{bc}^{-1} } \cup \mathbb Z^{3}_{{\rm det}[I_{ab}] I_{ab}^{-1} }). \nonumber
\end{eqnarray}
Furthermore  the following identity holds:
\begin{eqnarray}
T^{-t}\left(\begin{array}{c}
             \vec{\tilde{x}}\,[ \vec{\tilde{y}}]\\ \vec{\tilde{x}}\, [\vec{\tilde{y}}] \end{array}\right)=\left(\begin{array}{c}
             \vec{\tilde{x}}\,[\vec{\tilde{y}}]\\ 0\end{array}\right) \,\, .  \nonumber
\end{eqnarray}
After collecting all the results,  one can write:
\begin{eqnarray}
\phi_{\vec{j}_1}^{ab}(\Omega_{ab})\, \phi_{\vec{j}_2}^{bc}(\Omega_{bc})&=&{\cal N}_{ab}\,{\cal N}_{bc}\,e^{i\pi \vec{\tilde x}^t\,(I_{ab}+I_{bc}) \, \vec{\tilde y}+i\pi  \vec{\tilde y}^t (I_{ab}\,\Omega_{ab}+ I_{bc}\,\Omega_{bc}) \vec{\tilde y}}\!\!\!\!
\nonumber\\
  & =&   \! \! \frac{1}{\chi} \! \! \! \! \! \! \! \!  \sum_{\substack{
(\vec{l}_3,\,\vec{l}_4)\in \mathbb{Z}^{3} \\
\vec{ m}\in \tilde{\mathbb{Z}}_{(I_{ab}^{-1}+I_{bc}^{-1})\alpha}}}\sum_{\substack{\vec{p}\in\mathbb Z^{3}_{{\rm det}[I_{bc}] I_{bc}^{-1} }\\ \vec{\tilde p}\in  \mathbb Z^{3}_{{\rm det}[I_{ab}] I_{ab}^{-1} }  }}\!\!\!\!e^{i\pi \vec{l}^{~'t} \, Q' \, \vec{l}^{~'}+2\pi i \vec{l}^{~'t} \, Q' \, \left(\begin{array}{c}\vec{\tilde y}\\0\end{array}\right)
+ 2\pi i \vec{l}^{~'t} \, {\cal I}' \, \left(\begin{array}{c}\vec{\tilde x}\\0\end{array}\right)} \label{phiphi}
\end{eqnarray}
where the quantity
\begin{eqnarray}
\chi=\left\{\begin{array}{cc}
 |{\rm det}[I_{ab}]|^{d-1} &   {\rm if}\, I_{bc}=PI_{ab}\\
|{\rm det}[I_{bc}]|^{d-1}&  {\rm if}\, I_{ab}=PI_{bc}  \,\, \\ \,\,
1 &  {\rm independent } \,(I_{bc},I_{ab}) \,\, . \end{array}\right.\label{chi}
\end{eqnarray}
has been introduced.

The overlap of three wave functions results to be:
\begin{eqnarray}
&&\int_{0}^{1} d^3\tilde{x} d^3\tilde{y}\sqrt{G_6} \phi_{\vec{j}_3}^{ca}(\Omega_{ca})
\phi_{\vec{j}_1}^{ab}(\Omega_{ab})\, \phi_{\vec{j}_2}^{bc}(\Omega_{bc})={\cal N}_{ab}\,{\cal N}_{bc}\,{\cal N}_{ca} \int_{0}^{1} d^3\tilde{x} d^3\tilde{y}\sqrt{G_6} e^{i\pi \vec{\tilde y}^t\,(I_{ca}+I_{ab}+I_{bc}) \, \vec{\tilde x}}\nonumber\\
&\times & e^{i\pi  \vec{\tilde y}^t (I_{ca}\Omega_{ca}+ I_{ab}\,\Omega_{ab}+ I_{bc}\,\Omega_{bc}) \vec{\tilde y}}\frac{1}{\chi}\sum_{\substack{
(\vec{n}_3,\,\vec{l}_3,\,,\,\vec{l}_4)\in \mathbb{Z}^{3} \\
\vec{ m}\in  \tilde{\mathbb{Z}}_{(I_{ab}^{-1}+I_{bc}^{-1})\alpha}}}\sum_{\substack{\vec{p}\in\mathbb Z^{3}_{{\rm det}[I_{bc}] I_{bc}^{-1} }\\ \vec{\tilde p}\in  \mathbb Z^{3}_{{\rm det}[I_{ab}] I_{ab}^{-1} }  }}
e^{i\pi (\vec{n}_3+\vec{j}_3)^t  I_{ca}\, \Omega_{ca}(\vec{n}_3+\vec{j}_3)+ 2\pi i (\vec{n}_3+\vec{j}_3)^t I_{ca} (\vec{\tilde x}+\Omega_{ca}\vec{\tilde y})}\nonumber\\
&\times&e^{i\pi \vec{l}^{~'t} \,Q'\,  \vec{l}^{~'}+2\pi i \vec{l}^{~'t}\,  Q'\,  \left(\begin{array}{c}\vec{\tilde y}\\0\end{array}\right)
+ 2\pi i \vec{l}^{~'t}  {\cal I}'  \left(\begin{array}{c}\vec{\tilde x}\\0\end{array}\right)} \,\, .
\label{inteq}
\end{eqnarray}
The integral over the $\tilde{x}$ coordinates can be easily performed after using the identity \[ I_{ab}+I_{bc}+I_{ca}=0 \] and is given by:
\begin{eqnarray}
\int_0^1 d^3\tilde{x} e^{2\pi i ((\vec{n}_3+\vec{j}_3)^{t} I_{ca} +\vec{l}_1^{~'t}(I_{ab} +I_{bc} ))\vec{\tilde{x}}}
=\delta_{\vec{n}_3+\vec{j}_3, \vec{l}_1'}^{(3)}  \, .  \nonumber
\end{eqnarray}
This condition implies $\vec{n}_3=\vec{l}_3$ and
\begin{eqnarray}
(\vec{j}_1^t\,I_{ab}+\vec{j}_2^t\,I_{bc}+\vec{m}^t\,I_{ab})(I_{ab}+I_{bc})^{-1}= \vec{j}_3^t\label{delta}
\end{eqnarray}
which, when $I_{ab}$ and $I_{bc}$ are completely independent, gives  a well-defined condition on $\vec{m}$ because the numbers of inequivalent  $\vec{ m}$ and $\vec{j}$ coincide.
In the case in which $I_{ab}$ and $I_{bc}$ are not independent, the number of inequivalent values of $\vec{m}$ is smaller than the number of possible $\vec{j}$ (equal to  $|{\rm det}(I_{ab}+I_{bc})|$), and their degeneracy is $|{\rm det}  I_{ab}|^d$ or $|{\rm det}  I_{bc}|^d$ depending on the relation existing between the Chern classes. It follows that for a  given a value of $\vec{j}$ it is not possible, in general, to find an integer $\vec{m}$ satisfying eq. (\ref{delta}) and, when this is possible, it appears a number of times equal to the degeneracy of $\vec{m}$. The degeneracy can be taken into account by introducing:
\begin{eqnarray}
{\cal D}\equiv\frac{1}{\chi}\sum_{\vec{m}\in\tilde{\mathbb{Z}}_{(I_{ab}^{-1}+I_{bc}^{-1})\alpha}}
\delta_{(\vec{j}_1^t\,I_{ab}+\vec{j}_2^t\,I_{bc}+\vec{m}^t\,I_{ab})(I_{ab}+I_{bc})^{-1}; \vec{j}_3^t} \nonumber
\end{eqnarray}
and replacing $\vec{m}$,  in eq. (\ref{inteq}), with  the corresponding value of $\vec{j}_3$ as given by the identity (\ref{delta}).

In order to compute the integral over the variable $\tilde{y}$ one defines:
\begin{eqnarray}
A\equiv  I_{ca}\,\Omega_{ca}+Q'^{11}=I_{ca}\,\Omega_{ca} +I_{ab}\,\Omega_{ab} + I_{bc}\,\Omega_{bc}~~
\nonumber
\end{eqnarray}
with $A^t=A$, getting:
\begin{eqnarray}
&&\int d^3\tilde{y} e^{i\pi \vec{\tilde{y}}^{t} A \vec{\tilde{y}} +2\pi i {\vec{l}_1}^{~'t}  A \vec{\tilde{y}} +2\pi i {\vec{l}_2}^{~'t}  Q'^{21}\, \vec{\tilde{y}}}=\int d^3\tilde{y} e^{-\pi[\vec{\tilde{y}}^{t} +{\vec{l}_1}^{~'t}+{\vec{l}_2}^{~'t}\, Q'^{21}\,A^{-1}](-iA)[\vec{\tilde{y}}+\vec{l}^{~'}_1+A^{-1} {Q'^{21}}^t\vec{l}^{~'}_2]}\nonumber\\
&&\times e^{-i \pi [\vec{l}_1^{~'t}+\vec{l}_2^{~'t}\,Q'^{21}\,A^{-1}]A[\vec{l}_1^{~'}+A^{-1}\,{Q'^{21}}^t\, \vec{l}^{~'}_2]}
\equiv{\cal F}_{\Omega,I}(\vec{l}_3,\vec{l}_4)\,e^{-i \pi [\vec{l}_1^{~'t}+\vec{l}_2^{~'t}\,Q'^{21}\,A^{-1}]A[\vec{l}'_1+A^{-1}\,{Q'^{21}}^{t}\, \vec{l}^{~'}_2]}\nonumber
\end{eqnarray}
with
\begin{eqnarray}
{\cal F}_{\Omega,I}(\vec{l}_3,\vec{l}_4)\equiv \int_0^1 d^3\vec{\tilde{y}} e^{-\pi[\vec{\tilde{y}}^t +\vec{l}^{~'t}_1+\vec{l}^{~'t}_2\, {Q'^{21}}^{t}\,A^{-1}](-iA)[\vec{\tilde{y}}+\vec{l}^{~'}_1+A^{-1} {Q'^{21}}^t\vec{l}^{~'}_2]} \, .\label{int}
\end{eqnarray}
The Yukawa coupling becomes:
\begin{eqnarray}
&& \hspace{-1cm} \int d^3\tilde{x} d^3\tilde{y}\sqrt{G_6} {\phi_{\vec{j}_1}^{ca}(\Omega_{ca})}^*
\phi_{\vec{j}_1}^{ab}(\Omega_{ab})\, \phi_{\vec{j}_2}^{bc}(\Omega_{bc}) ={\cal N}_{ab}\,{\cal N}_{bc}\,{\cal N}_{ca}\sqrt{G_6}{\cal D} \sum_{\substack{
\vec{l}_3,\vec{l}_4\in \mathbb{Z}^{3}}}\sum_{\substack{\vec{p}\in\mathbb Z^{3}_{{\rm det}[I_{bc}] I_{bc}^{-1} }\\ \vec{\tilde p}\in  \mathbb Z^{3}_{{\rm det}[I_{ab}] I_{ab}^{-1} }  }} {\cal F}_{\Omega,I}(\vec{l}_3,\vec{l}_4) \nonumber \\
&&  \hspace{-1cm} \times  e^{i\pi \vec{l}^{~'t}_1[ I_{ca}\Omega_{ca}+ Q'^{11}- A]\vec{l}^{~'}_1+i\pi \vec{l}^{~'}t_1[Q'^{12}-{Q'^{21}}^t] \vec{l}_2^{~'}+
i\pi \vec{l}^{~'t}_2[Q'^{21}-Q'^{21}] \vec{l'}_1+i\pi \vec{l}_2^{~'t}[Q'^{22}-Q'^{21}A^{-1}{Q^{21}}^t] \vec{l}^{~'}_2}\nonumber\\
&&\hspace{-1cm} =\sqrt{G_6}  {\cal D}{\cal N}_{ab}\,{\cal N}_{bc}\,{\cal N}_{ca} \sum_{\substack{
\vec{l}_3,\vec{l}_4\in \mathbb{Z}^{3}}}\sum_{\substack{\vec{p}\in\mathbb Z^{3}_{{\rm det}[I_{bc}] I_{bc}^{-1} }\\ \vec{\tilde p}\in  \mathbb Z^{3}_{{\rm det}[I_{ab}] I_{ab}^{-1} }  }}{\cal F}_{\Omega,I}(\vec{l}_3,\vec{l}_4)
e^{i\pi \vec{l}_2^{'t}~ \Pi~ \vec{l}^{~'}_2} \nonumber
\end{eqnarray}
where  eq. (\ref{delta}) has been used and
\begin{eqnarray*}
\Pi=\alpha\left((\Omega_{ab}I_{ab}^{-t}+\Omega_{bc}I_{bc}^{-t})-(\Omega_{ab}-\Omega_{bc})
(I_{ca} \Omega_{ca}+I_{ab} \Omega_{ab}+I_{bc}\Omega_{bc})^{-1}(\Omega_{ab}-\Omega_{bc})^t\right)
\alpha^{-t} \,\, . \nonumber
\end{eqnarray*}
The integral in eq. (\ref{int}) is convergent and can be  explicitly computed. The key ingredient in order to prove the convergence of the integral is the
inequality:
\begin{eqnarray}
&&\int_0^1 d^3 {\tilde{y}}
 \sum_{ \vec{l}_3\in \mathbb{Z}^d}e^{-\pi[ \vec{\tilde{{y}}}^t +\vec{l}_1^{~'t}+{\vec{l}_2}^{~'t}\,Q'^{21}\,A^{-1}](-iA)[\vec{\tilde{y}}+\vec{l}'_1+A^{-1}{Q'^{21}}^t
 \vec{l}^{~'}_2]}\nonumber\\
&&\leq
\int d^3 {\tilde{y}} \sum_{\vec{l}_3\in \mathbb{Z}^d}\left|e^{-\pi[ \vec{\tilde{y}}
^t +{\vec{l}_1}
^{~'t}+\vec{l}_2^{~'t}\, Q'^{21}\,A^{-1}](-iA)[\vec{\tilde{y}}+\vec{l}^{~'}_1+A^{-1}{Q'^{21}}^t\vec{l}^{~'}_2]}\right|\nonumber\\
&=&e^{\pi[\vec{l'}_2 {\rm Im}(Q'^{21}
A^{-1})[{\rm Re}A\,{\rm Im}A^{-1}\,{\rm Re A}- {\rm Im}A]{\rm Im}(Q'^{21}A^{-1})^t\vec{l}_2^{~'}]} \sum_{\vec{l}_3\in \mathbb{Z}^d}\int_0^1 d^3 {\tilde{y}}\nonumber\\
&\times&e^{-\pi [\vec{\tilde{y}}^t +\vec{l}_1^{~'t}+\vec{l}_2^{~'}\,{\rm Re}(Q'^{21}A^{-1})+\vec{l}_2^{~'t}{\rm Im}(Q'^{21}A^{-1}){\rm Re}A\,{\rm Im}A^{-1}]{\rm Im}A[ \vec{\tilde{y}}+ \vec{l}_1^{~'t}+ {\rm Re}(Q'^{21}A^{-1})^t\vec{l}^{~'}_2+ {\rm Im}A^{-1}\,{\rm Re}A {\rm Im}( Q'^{21}A^{-1})^t\vec{l}_2^{~'}]}\nonumber\\
&\equiv&e^{\pi[\vec{l}^{~'}_2 {\rm Im}(Q'^{21}A^{-1})[{\rm Re}A\,{\rm Im}A^{-1}\,{\rm Re A}- {\rm Im}A]{\rm Im}(Q'^{21}A^{-1})^t\vec{l}^{~'}_2]} \sum_{\vec{l}^{~'}_3\in \mathbb{Z}^d}
\int_{0}^{1}d^3 {\tilde{y}} f(\vec{\tilde{y}}+\vec{l}_3+\vec{j}_3+\vec{\hat{l}}_2^{~'})   \nonumber
\end{eqnarray}
with
\begin{eqnarray}
&&f(\vec{\tilde{y}}+\vec{l}_3+\vec{j}_3+\vec{\hat{l}}_2^{~'})=e^{-\pi[ \vec{\tilde{y}}+\vec{l}_3+\vec{j}_3+\vec{\hat{l}}_2^{~'}]^t\,{\rm Im}A\,[ \tilde{y}+\vec{l}_3+\vec{j}_3+\vec{\hat{l}}_2^{~'}]}\nonumber\\
&&\vec{\hat{l}}_2^{~'}= \left[{\rm Re}(Q'^{21}A^{-1})^t+ {\rm Im}A^{-1}\,{\rm Re}A {\rm Im}( Q'^{21}A^{-1})^t\right]\vec{l}_2^{~'} \,\, .\nonumber
\end{eqnarray}
By observing that
\begin{eqnarray}
\lim_{\vec{\delta}\rightarrow \infty}\sum_{\vec{l}_3=\vec{-\delta}}^{+\vec{\delta}-\vec{1}}
\int_{0}^{1}d^3{\tilde{y}} f(\vec{\tilde{y}}+\vec{l}_3+\vec{j}_3+\vec{\hat{l}}_2^{~'})=\lim_{\vec{\delta}\rightarrow \infty}\int_{-\vec{\delta}+\vec{j}_3+\vec{\hat{l}}_2^{~'}}^{\vec{\delta}+\vec{j}_3+\vec{\hat{l}}_2^{~'}}d^3
{\tilde{y}} f(\vec{\tilde{y}})= \int_{-\infty}^{\infty}d^3{\tilde{y}} e^{-\pi \vec{\tilde{y}}^t\,{\rm Im}A\,\vec{\tilde{y}}} \nonumber\label{162}
\end{eqnarray}
one sees that the integral is finite because ${\rm Im}A$ is positive definite.

After having proved the convergence of this integral,  one can now explicitly compute it.  One can introduce the complex variable
\begin{eqnarray*}
w^i&=&{\tilde y}^i+\vec{l}_3^i+\vec{j}^i +[{\rm Re}(Q'^{21}A^{-1})^t \vec{l}_2^{~'}]^i +i [{\rm Im}( Q'^{21}A^{-1})^t\vec{l}^{~'}_2)]^i \nonumber
\end{eqnarray*}
and the integral becomes
\begin{eqnarray}
&&\prod_{i=1}^3\left[\lim_{\delta^i\rightarrow \infty} \sum_{\vec{l}_3^i=-{\delta^i}}^{\delta^i -1} \int_{ {l}_3^i+{j}_3^i +[{\rm Re}(Q'^{21}A^{-1})^t \vec{l}_2^{~'}]^i +i [{\rm Im}( Q'^{21}A^{-1})^t\vec{l}^{~'}_2]^i }^{{l}_3^i+{j}^i +[{\rm Re}(Q'^{21}A^{-1})^t \vec{l}_2^{~'}]^i +i [{\rm Im}( Q'^{21}A^{-1})^t\vec{l}^{~'}_2]^i +1} d w^i\right] e^{- \pi \vec{w}^t \,(-i A )\, \vec{w}}\nonumber\\
&&=\prod_{i=1}^3\left[ \lim_{\delta^i\rightarrow \infty}\int_{-\delta^i +i ({\rm Im}( Q'^{21}A^{-1})^t\vec{l}_2')^i}^{\delta^i +i ({\rm Im}( Q'^{21}A^{-1})^t\vec{l}_2')^i}d w^i\right] e^{- \pi \vec{w}^t \,(-i A )\, \vec{w}}=\int_{-\infty}^{+\infty} d^3 w e^{-\pi \vec{w}^t \,(-i A )\, \vec{w}}\nonumber\\
&&= ({\rm det} (-iA))^{-1/2}  \nonumber
\end{eqnarray}
In the last step of the previous equation,  being the integrand an analytic function of each of its variables,   it is possible to apply the Cauchy theorem.
Consequently, one can write:
\begin{eqnarray}
0& = & \oint d^3w e^{- \pi \vec{w}^t \,(-i A )\, \vec{w}} = \prod_{i=1}^{3} \left[ \int_{+\delta_{i}}^{-\delta_{i}} + \int^{-\delta^i +i ({\rm Im}( Q'^{21}A^{-1})^t\vec{l}_2')^i}_{-\delta^i}  \right. \nonumber \\
&& + \left.  \int_{-\delta^i +i ({\rm Im}( Q'^{21}A^{-1})^t\vec{l}_2')^i}^{\delta^i +i ({\rm Im}( Q'^{21}A^{-1})^t\vec{l}_2')^i} + \int_{\delta^i +i ({\rm Im}( Q'^{21}A^{-1})^t\vec{l}_2')^i}^{\delta^{i}}  \right] d^{3} w \,\, e^{- \pi \vec{w}^t \,(-i A )\, \vec{w}}   \,\, . \nonumber
\end{eqnarray}
The integrals along the directions parallel to the imaginary axis are weighted by the factor $e^{-\pi \vec{\delta}\,  \mbox{Im} A \, \vec{\delta}}$ that vanishes in the limit $\vec{\delta} \rightarrow \infty$. This proves the last equality in eq. (\ref{162}).
By collecting all the results, one gets the following expression for the Yukawa couplings:
\begin{eqnarray}
&&\int d^3\tilde{x} d^3\tilde{y}\sqrt{G_6} {\phi_{\vec{j}_1}^{ca}(\Omega_{ca})}^*
\phi_{\vec{j}_1}^{ab}(\Omega_{ab})\, \phi_{\vec{j}_2}^{bc}(\Omega_{bc}) = \sqrt{G_6} {\cal C}_{ab} {\cal C}_{bc} {\cal C}_{ca} (\mbox{det} (-iA))^{-1/2} \nonumber\\
&& \,\times {\cal D} \sum_{\substack{\vec{p}\in\mathbb Z^{3}_{{\rm det}[I_{bc}] I_{bc}^{-1} }\\ \vec{\tilde p}\in  \mathbb Z^{3}_{{\rm det}[I_{ab}] I_{ab}^{-1} }  }}
\Theta\left[\substack{  \frac{I_{bc}^t}{{\rm det }[I_{ab}I_{bc}]}(\vec{j}_3-\vec{j}_2) +\frac{I_{bc}^t}{{\rm det}I_{bc}}\vec{p}+ \frac{I_{ab}^t}{{\rm det}I_{ab}}\vec{\tilde p}
\\0}\right](0| \Pi) \,\, .  \nonumber
\end{eqnarray}

In the last part of this appendix we would like to give some more comments about the case in which $I_{ab}$ and $I_{bc}$ commute. In this case the quantities
$\alpha (I_{ab}^{-1}+I_{bc}^{-1})$ are integer matrices with the choice $\alpha= I_{ab}I_{bc}$ and  eqs. ({\ref{mdef}) become:
\begin{eqnarray}
&&\left(\vec{n}_1^t \,I_{ab}+\vec{n}_2^t\,I_{bc}\right)\left(I_{ab}+I_{bc}\right)^{-1}= \vec{m}_1^t \left(I_{ab} +I_{bc} \right)^{-1}+\vec{l}_3^t \nonumber \\
&& \left( \vec{n}_1^t-\vec{n}_2^t \right) \left(I_{ab}+I_{bc}\right)^{-1} = \vec{m}_2^t \left(I_{ab} +I_{bc} \right)^{-1} +\vec{l}_4^t  .  \nonumber
\end{eqnarray}
Now the sets of the integer values taken by $\vec{m}_{1}$ and $\vec{m}_{2}$ coincide and there is no need to introduce the vectors $\vec{p}$ and $\vec{\tilde{p}}$, which leads to a simplification  in the sum appearing in  eq. (\ref{phiphi}).

The identity written in eq. (\ref{yukcom}) is proved starting from the indentity
\begin{eqnarray}
\vec{t}^t\left[\frac{I_{ab}}{{\rm det}[I_{ab}]}\right] \left[\frac{I_{bc}}{{\rm det}[I_{bc}]}\right]
= \vec{n}^t \left[\frac{I_{ab}}{{\rm det}[I_{ab}]}\right] \left[\frac{I_{bc}}{{\rm det}[I_{bc}]}\right]+\vec{l}~~\label{ehsan};
\end{eqnarray}
valid for $\vec{l}$ and $\vec{t}$  arbitrary integers and $\vec{n}^t$ defined up the  translation:
\[ \vec{n}^t\equiv \vec{n}^t+ \vec{k}^t \left[\frac{I_{ab}}{{\rm det}[I_{ab}]}\right]^{-1} \left[\frac{I_{bc}}{{\rm det}[I_{bc}]}\right]^{-1}.\]
 Eq. (\ref{ehsan}) can be understood by observing that, being $\vec{t}$ an arbitrary integer,  when it is proportional to $\left[\frac{I_{ab}}{{\rm det}[I_{ab}]}\right]^{-1} \left[\frac{I_{bc}}{{\rm det}[I_{bc}]}\right]^{-1}$ one gets the last terms on the right of the identity, otherwise one gets  the first term on the right. The number of non equivalents $\vec{n}$ is $|{\rm det} [\left[\frac{I_{ab}}{{\rm det}[I_{ab}]}\right]^{-1} \left[\frac{I_{bc}}{{\rm det}[I_{bc}]}\right]^{-1}]|$. Furthermore, when $I_{ab}$ and $I_{bc}$ are completely independent (see comment before eq. (\ref{yukt6})), it is possible to write:
\begin{eqnarray}
\vec{t}^t\left[\frac{I_{ab}}{{\rm det}[I_{ab}]}\right] \left[\frac{I_{bc}}{{\rm det}[I_{bc}]}\right]
= \vec{p}^t \left[\frac{I_{ab}}{{\rm det}[I_{ab}]}\right]+\vec{\tilde p}^t \left[\frac{I_{ab}}{{\rm det}[I_{bc}]}\right]+\vec{l}^t~~   \nonumber
\end{eqnarray}
or, equivalently:
\begin{eqnarray}
\vec{t}^t
= \vec{p}^t \left[\frac{I_{bc}}{{\rm det}[I_{bc}]}\right]^{-1}+\vec{\tilde p}^t \left[\frac{I_{ab}}{{\rm det}[I_{ab}]}\right]^{-1}+\vec{l}^t \left[\frac{I_{ab}}{{\rm det}[I_{ab}]}\right]^{-1} \left[\frac{I_{bc}}{{\rm det}[I_{bc}]}\right]^{-1}.    \nonumber
\end{eqnarray}
By using this latter identity, the left side of eq. (\ref{yukcom}) can be alternatively written as:
\begin{eqnarray}
&&\sum_{\substack{\vec{p}\in\mathbb Z^{3}_{{\rm det}[I_{bc}] I_{bc}^{-1} }\\ \vec{\tilde p}\in  \mathbb Z^{3}_{{\rm det}[I_{ab}] I_{ab}^{-1} }  }}\sum_{\vec{l}\in \mathbb{Z}^3} e^{-i\pi \left[P^t\vec{l}+ \frac{P^tI_{bc}^t}{{\rm det} I_{ab}I_{bc}}(\vec{j}_3-\vec{j}_2)+\frac{P^tI_{bc}}{{\rm det} I_{bc}^t}\vec{p}+\frac{P^tI_{ab}^t}{{\rm det}I_{bc}}\vec{\tilde{p}}\right]^t\tilde{\Pi}\left[P^t\vec{l}+ \frac{P^tI_{bc}^t}{{\rm det} I_{ab}I_{bc}}(\vec{j}_3-\vec{j}_2)+\frac{P^tI_{bc}}{{\rm det} I_{bc}^t}\vec{p}+\frac{P^tI_{ab}^t}{{\rm det}I_{bc}}\vec{\tilde{p}}\right]}\nonumber\\
&=& \sum_{\vec{t}\in \mathbb{Z}^3}e^{-i\pi \left[t+ I_{ab}^{-t}(\vec{j}_3-\vec{j}_2)\right]^t\,\tilde{\Pi}\,
\left[t+ I_{ab}^{-t}(\vec{j}_3-\vec{j}_2)\right]}  \nonumber
\end{eqnarray}
which proves eq. (\ref{yukcom}), after reminding the definition of $P={\rm det} [I_{ab}I_{bc }]I_{ab}^{-1}I_{bc}^{-1}$.

\section{Strings in a magnetic background} \label{string}
In the string approach to the toroidal compactification, a $2d$-real torus is the  lattice in the plane $\mathbb{R}^{2d}$ made by a collection of $2d$ vectors $\vec{e}_M$ ($M=1,\dots, 2d$)  together with the identification
\begin{eqnarray}
\vec{x}\equiv \vec{x}+2\pi\ \sqrt{\alpha'} \,m^M\, \vec{e}_M~~,~~\vec{x}\in \mathbb{R}^{2d}~~\mbox{and}~~\vec{m}\in \mathbb{Z}  \,\,\, .  \nonumber
\end{eqnarray}
In string theory, the parameter $R$,  defined in eq. (\ref{ident}), is identified with  the string slope.
In a Cartesian frame the components of the defining lattice vectors are $\vec{a}_M\equiv (a_{~M}^I)$ and the metric is $G_{MN}= e^{J}_{~N}\delta_{IJ}e^I_{~M}$, and by construction $e^I_{~M}$ is the vielbein of the metric. Any other matrix of the kind $E=O\,e$ obtained by an orthogonal rotation of $e$ is a good vielbein matrix. The complex coordinates are introduced through the complex vielbein:
\begin{eqnarray}{\cal E}=S\,E ~~,~~S=\frac{1}{\sqrt{2}}\left(\begin{array}{cc}
                                                             1&i\\
                                                             1&-i\end{array}\right) \,\, ,    \nonumber
\end{eqnarray}
being $S$ the matrix which diagonalizes the complex structure $J_{\mathbb{R}} = \left(\begin{array}{cc} 0&-1\\
                                                             1&0\end{array}\right)$:

\begin{eqnarray}
J_{\mathbb{C}} \equiv \left(\begin{array}{cc}
                                                             i&0\\
                                                             0&-i\end{array}\right)=S\,J_{\mathbb{R}}\,S^{-1} .
                              \nonumber                               \end{eqnarray}
Here, a simplified notation is used where all the four blocks of the matrices are proportional to the identity matrix. In the complex coordinates, the flat metric ${\cal G}$ is off-diagonal:
\begin{eqnarray}
{\cal G}={\cal E}^{-T} G {\cal E}^{-1}  \,\, .  \nonumber
\end{eqnarray}
The antisymmetric two-form ${\cal F}= B+2\pi \alpha' F$ in the Cartesian frame
\begin{eqnarray}
{\cal F}= E^{-T}{\cal F} E^{-1}= EG^{-1} {\cal F} E^{-1} \,\, ,  \nonumber
\end{eqnarray}
being $B$ the Kalb-Ramond field,  can be reduced by an orthogonal transformation $O_f$ in the block diagonal form \cite{0709.1805}:
\begin{eqnarray}
{\cal F}_{bd}=\left( \begin{array}{cc}
                        0&\mathbb{I}_{\xi}\\
                        -\mathbb{I}_{\xi}&0\end{array}\right)=O_f {\cal F}_c O_f^{T}=O_f E{\cal F}  E^{-1}O_f^{-1}= {E'}_f{\cal F}{E'}_f^{-1}   \nonumber
\end{eqnarray}
with $E'_{f}=O_f\,E$ and $\mathbb{I}_\lambda={\rm diag}(\lambda_1,\dots \lambda_d)$ with $\lambda\in \mathbb{R}$. The vielbein $E_f$ transforms the metric into the identity and the antisymmetric field  into a block-diagonal matrix. This kind of vielbein can also be used to introduce a particular set of coordinates (${\cal E}_f= SE_f$)
which diagonalize the matrix
\begin{eqnarray}
{\cal F}^d={\cal E}_fG^{-1}{\cal F}{\cal E}_f^{-1}= {\cal G}E^{-T}{\cal F}{\cal E}^{-1}=\left( \begin{array}{cc}
-i\mathbb{I}_{\lambda}&0\\
0&i\mathbb{I}_{\lambda}\end{array}\right)  \,\, . \label{bdf}
\end{eqnarray}
The boundary conditions of an open string ending on two branes with different magnetization depend, in the bosonic sector, on the monodromy matrix $R=R^{-1}_\pi R_0$ with:
\begin{eqnarray}
R_\sigma=(G-{\cal F}_\sigma)^{-1}(G+{\cal F}_\sigma)=
\left(\mathbb{I}-G^{-1}{\cal F_{\sigma} }\right)^{-1}\left( \mathbb{I}+G^{-1}{\cal F_{\sigma}}\right)~~;~~\sigma=0,\pi \,\, .  \nonumber
\end{eqnarray}
In the complex basis, above defined,  this operator is a diagonal matrix. However, we are interested in computing the Yukawa couplings involving open strings  ending on stacks of magnetized branes having different magnetic fields turned on in their world-volume. In this complex frame, only one of these fields can be recast in the block-diagonal form. In the following, all the monodromy matrices are assumed to commute. According to ref. \cite{0709.1805} this implies
 that all the magnetic fields can be taken as in eq. (\ref{bdf}).

The monodromy matrix is diagonal in the complex frame, being equal to
\begin{eqnarray}
{\cal E}_{f}\, R\, {\cal E}_{f}^{-1}= {\cal E}_{f}\, R^{-1}_\pi{\cal E}_{f}^{-1}{\cal E}_{f}R_0\, {\cal E}_{f}^{-1}    \nonumber
\end{eqnarray}
and then by evaluating
\begin{eqnarray}
{\cal E}_f\, R_\sigma\, {\cal E}^{-1}_{f}={\cal E}_f\,(\mathbb{I}- G^{-1}{\cal F}_{\sigma})^{-1} (\mathbb{I}+G^{-1}{\cal F}_{\sigma}) \, {\cal E}^{-1}_{f} \nonumber
\end{eqnarray}
it follows:
\begin{eqnarray}
&&{\cal E}_{f}\, R_\sigma\, {\cal E}_{f}^{-1}=\left(\mathbb{I}- {\cal E}_{f}  \, G^{-1}{\cal F}_\sigma\,{\cal E}_{f} ^{-1}\right)^{-1}(\mathbb{I}+{\cal E}_{f} G^{-1}{\cal F}_\sigma{\cal E}_{f}^{-1})\nonumber\\
&&={\rm diag}\left( \frac{1 +i\hat{\lambda}_1^\sigma}{1 -i\hat{\lambda}_1^\sigma},\dots ,\frac{1 +i\hat{\lambda}_{d}^\sigma}{1 -i\hat{\lambda}_{d}^\sigma},
\frac{1 -i\hat{\lambda}_1^\sigma}{1 +i\hat{\lambda}_1^\sigma},\dots ,\frac{1 -i\hat{\lambda}_{d}^\sigma}{1 +i\hat{\lambda}_{d}^\sigma}\right) \nonumber
\end{eqnarray}
By defining the eigenvalues of the matrix $R_\sigma=\left( e^{2\pi\nu_a^\sigma},\,e^{-2\pi\nu_a^\sigma}\right)$, one has:
\begin{eqnarray}
e^{2\pi\nu_a^\sigma}= \frac{1 +i\hat{\lambda}_a^\sigma}{1 -i\hat{\lambda}_a^\sigma}\Rightarrow \tan\pi\nu_a^\sigma= \hat{\lambda}^\sigma_a \nonumber
\end{eqnarray}
In the same way, one gets:
\begin{eqnarray}
{\cal E}_{f} \, R\, {\cal E}_{f}^{-1}= {\rm diag} \left( \frac{(1 -i\hat{\lambda}_1^\pi)(1 +i\hat{\lambda}_a^0)}{(1 +i\hat{\lambda}_1^\pi)(1 -i\hat{\lambda}_1^0)}\dots,\,\frac{(1 +i\hat{\lambda}_{a}^\pi)(1 -i\hat{\lambda}_{a}^0)}{(1 -i\hat{\lambda}_{1}^\pi)(1 +i\hat{\lambda}_{1}^0)}\dots\right)   \nonumber
\end{eqnarray}
In terms of its eigenvalues, this matrix  is  usually denoted by:
\[
 {\cal E}_{f}\, R\, {\cal E}_{f}^{-1}= {\rm diag} \left(e^{2\pi i\nu_a}\, , e^{-2\pi i\nu_a}\right)
 \]
 which leads to the identification:
\begin{eqnarray}
e^{2\pi i\nu_a}= \frac{(1 -i\hat{\lambda}_a^\pi)(1 +i\hat{\lambda}_a^0)}{(1 +i\hat{\lambda}_a^\pi)(1 -i\hat{\lambda}_a^0)} \,\, .   \nonumber
\end{eqnarray}
The previous identity determines:
\begin{eqnarray}
\tan \pi \nu_a=\frac{{\hat{\lambda}}_a^0 -{\hat{\lambda}}^\pi_a}{1 + {\hat{\lambda}}_a^0 {\hat{\lambda}}^\pi_a}=\frac{ \tan\pi\nu_a^0-\tan\pi\nu_a^\pi}{
1+ \tan\pi\nu_a^0 \tan\pi\nu_a^\pi}=\tan \pi(\nu_a^0-\nu_a^\pi) \,\, .\label{nulambda}
\end{eqnarray}
By setting $B=0$ and reminding that ${\cal F}=2\pi\alpha'F$,  one gets  the relation between the gauge field defined in the field theory approach
and  the corresponding stringy quantity:
\begin{eqnarray}
\frac{\hat{\lambda}}{2\pi \alpha'}= \frac{1}{(2\pi R)^2} \lambda   \nonumber
\end{eqnarray}
where Eqs. (\ref{89}) and (\ref{bdf}) have been used. By performing the zero-slope limit, keeping fixed the field theory quantities, $R$ and $\lambda$, the quantity $\hat{\lambda}$ is necessarily small. From eq. (\ref{nulambda}), in the field theory limit, one has $\tan\pi\nu_a \sim \pi\nu_a \sim \hat{\lambda}_a^{0} - \hat{\lambda}_a^{\pi}$.
The string Hamiltonian is \cite{0512067,1101.0120}:
\begin{eqnarray}
H =-\alpha'^2 p^2 + \left[  N^{X}  + N^{\psi} + \sum_{r=1}^{3} \left( N^{\cal{Z}}_{r}  +
N^{\Psi}_{r}
\right)  - \frac{x}{2}
+ \frac{x}{2} \sum_{a=1}^{3} |\nu_a|      \right]
\label{M2}
\end{eqnarray}
where $x=1\, (0)$ for the NS (R) sector, being the $N$s the number operators.

In the field theory limit, as explained in sec. 2.2 of ref \cite{1101.0120}, the mass formula reduces to:
\begin{eqnarray}
M_r^2= \sum_{r=1}^3 \frac{|\nu_r|}{2\alpha'}(2N_r+1) \pm \frac{|\nu^s|}{\alpha'}\sim \sum_{r=1}^3\frac{| \lambda_r | }{(2\pi R)^2}(2N_r+1) \pm \frac{2 \lambda_s}{(2 \pi R)^2}  \nonumber
\end{eqnarray}
which coincides with eq. (\ref{m2}).


\begin{thebibliography}{77}

 \bibitem{IU} L. E. Ib\`a\~nez and A. Uranga, {\em String Theory and Particle Physics. An Introduction to String Phenomenology,}  Cambridge University  Press (2012).

 \bibitem{Review1} R. Blumenhagen, B. K\"ors, D. L\"ust and S. Stieberger, {\em Four-dimensional String Compactifications with D-Branes, Orientifolds and Fluxes}, Phys.Rept. {\bf 445} (2007) 1, [{\tt hep-th/0610327}].
   \bibitem{Review2}   M. Cvetic and J. Halverson, {\em TASI Lectures: Particle Physics from Perturbative and Non-perturbative Effects in D-braneworlds}, [{\tt arXiv:1101.2907 [hep-th]}].
        \bibitem{Review3}L. Ibanez, \emph{From Strings to the LHC. Les Houches Lectures on String Phenomenology},  [{\tt arXiv:1204.5296 [hep-th]}].


\bibitem{magnetized1}
E. S. Fradkin and A. A. Tseytlin, \emph{Nonlinear Electrodynamics from Quantized Strings},
Phys. Lett. {\bf B163} (1985) 123.
\bibitem{magnetized2}
A. A. Tseytlin, \emph{Renormalization of M\"{o}bius Infinities and Partition Function Representation for
String Theory Effective Action}, Phys. Lett. {\bf B202} (1988) 81.
\bibitem{magnetized3}
A. Abouelsaood, C. G. Callan Jr., C. R. Nappi and S. A. Yost,\emph{ Open Strings in Background Gauge Fields}, Nucl. Phys. {\bf B280} (1987) 599.
\bibitem{magnetized4}
C. Bachas,  \emph{A way to break supersymmetry},  [{\tt hep-th/9503030}].


\bibitem{0512067}
M. Bertolini, M. Bill\`{o}, A. Lerda, J. F. Morales and Rodolfo Russo, \emph{Brane world effective actions for D-branes with fluxes},
Nucl. Phys. {\bf B743} (2006) 1, [{\tt hep-th/0512067}].
\bibitem{1201.3604}
G. Honecker and J. Vanhoof, \emph{	
Yukawa couplings and masses of non-chiral states for the Standard Model on D6-branes on T6/Z6'},  JHEP {\bf 1204} (2012) 085, [{\tt arXiv:1201.3604 [hep-th]}].

\bibitem{0508043}
M. Berg, M. Haack and B. Kors, \emph{String Loop Corrections to K\"ahler Potentials in Orientifolds}, JHEP {\bf 0511} (2005) 030	[{\tt hep-th/0508043}].
\bibitem{1112.5156}	
M. Berg, M. Haack and J. U. Kang,
\emph{One-Loop K\"ahler Metric of D-Branes at Angles}, [{\tt  arXiv:1112.5156 [hep-th]}].

\bibitem{0404229}
D. Cremades, L. E. Ibanez and F. Marchesano, \emph{Computing Yukawa
couplings from magnetized extra dimensions},
JHEP {\bf 0405} (2004) 079,
[{\tt hep-th/0404229}].

\bibitem{also1}
J. P. Conlon, A. Maharana and F. Quevedo, \emph{Wave Functions and Yukawa Couplings in Local String Compactifications}, JHEP {\bf 0809} (2008) 104,
[{\tt arXiv:0807.0789  [hep-th]}].
\bibitem{also2}
F. Marchesano, P. McGuirk and  G. Shiu, \emph{Open String Wavefunctions in Warped Compactifications}, JHEP {\bf 0904} (2009) 095,
[{\tt arXiv:0812.2247 [hep-th]}].
\bibitem{also3}
H. Abe, K. S. Choi, T. Kobayashi and  H. Ohki, \emph{Three generation magnetized orbifold models} Nucl. Phys. {\bf B814} (2009) 265,
[{\tt arXiv:0812.3534 [hep-th]]}.
\bibitem{also4}
P. Di Vecchia, A. Liccardo, R. Marotta and F. Pezzella, {\em K\"ahler Metrics: String vs Field Theoretical Approach},    Fortsch.Phys. {\bf 57} (2009) 718,  [{\tt arXiv:0901.4458v1 [hep-th]]}.
\bibitem{also5}
 H. Abe, K. S. Choi, T. Kobayashi and  H. Ohki, \emph{
    Higher Order Couplings in Magnetized Brane Models
    Hiroyuki Abe, Kang-Sin Choi, Tatsuo Kobayashi, Hiroshi Ohki} JHEP {\bf 0906} (2009) 080,
    [{\tt  arXiv:0903.3800  [hep-th]}].
   \bibitem{also6}
    H. Abe, T. Kobayashi, H. Ohki and K. Sumita, \emph{     	
Superfield description of 10D SYM theory with magnetized extra dimensions}, [{\tt arXiv:1204.5327 [hep-th]}].

    \bibitem{0709.4149}
P. Di Vecchia, A. Liccardo, R. Marotta, I. Pesando and  F. Pezzella, \emph{Wrapped Magnetized Branes: Two Alternative Descriptions?}, JHEP {\bf 0711} (2007) 100,
[{\tt arXiv:0709.4149 [hep-th]}].
\bibitem{0906.3033}
 P. G. C\'amara and F. Marchesano, {\em Open string wavefunction in flux compactification},  JHEP 0910 (2009) 017,  [{\tt arXiv:0906.3033 [hep-th]}].

\bibitem{0904.0910}
I. Antoniadis, A. Kumar and B. Panda, { \em Fermion Wavefunctions in Magnetized branes: Theta identities and Yukawa couplings}, Nucl. Phys. {\bf B823} (2009) 116, [{\tt arXiv:0904.0910 [hep-th]}].
\bibitem{MumfordI} D. Mumford, {\em Tata Lectures
on Theta I}, Birkh\"auser, Boston 1983.

\bibitem{0709.1805} D. Duo, R. Russo and S. Sciuto, {\em  New twist field couplings from the partition function for multiply wrapped D-branes}, JHEP {\bf 0712} (2007) 042, [{\tt arXiv:0709.1805 [hep-th]}].

\bibitem{BachasPorrati} C. Bachas and M. Porrati, {\em Pair creation of open strings in an electric fields}, Phys. Lett. {\bf B296} (1992) 77, [{\tt hep-th/9909032}];

\bibitem{GSW} M. Green, J. H. Schwarz and E. Witten, {\em Superstring Theory}, Vol. II, Cambridge University Press (1987).



\bibitem{0810.5509} P. Di Vecchia, A. Liccardo, R. Marotta and F. Pezzella, \emph{K\"{a}hler Metrics and Yukawa Couplings in Magnetized Brane Models}, JHEP {\bf 0903} (2009) 029,
[{\tt arXiv:0810.5509  [hep-th]}].
\bibitem{1110.5359} P. Anastasopoulos, M. Bianchi and R. Richter, \emph{On Closed-String Twist-Field Correlators and their Open-String Descendants}, [{\tt arXiv:1110.5359 [hep-th]}].


\bibitem{THooft} G. W. 't Hooft, {\em Some twisted self-dual solutions for the Yang-Mills equations
on a hypertorus}, Commun. Math. Phys. {\bf 81} (1981) 267.
\bibitem{0306006} 	M. Sakamoto and S. Tanimura, {\em An Extension of Fourier analysis for the n torus in the magnetic field and its application to spectral analysis of the magnetic Laplacian},  J. Math. Phys. {\bf 44} (2003) 5042, [{\tt hep-th/0306006}].
\bibitem{0701292}
R. Russo and S. Sciuto, 
\emph{The twisted open string partition function and Yukawa couplings},
 JHEP {\bf 0704} (2007) 030,  [{\tt hep-th/0701292}].

\bibitem{complex}
C. Birkenhake and H. Lange, {\em Complex Tori},  Progress in Mathematics,  vol. 177.


\bibitem{1101.5898} I. Pesando, \emph{Strings in an arbitrary constant magnetic field with arbitrary constant metric and stringy form factors},  [{\tt arXiv:1101.5898 [hep-th]}].

\bibitem{1101.0120}P. Di Vecchia, R. Marotta, I. Pesando and F. Pezzella, {\em Open strings in the system D5/D9}, J. Phys.  {\bf A44} (2011), 245401, [{\tt arXiv:1101.0120 [hep-th]}].





\bibitem{Nakahara} M. Nakahara, {\em Geometry, Topology and Physics}, Adam Hilger, Bristol and New York.
\bibitem{0804.0213}  C. Mantz and T. Prokopec, {\em Hermitian Gravity and Cosmology},  [{\tt arXiv:0804.0213 [gr-qc]}].






\end{thebibliography}
\end{document}